\documentclass{elsarticle}

\usepackage[utf8]{inputenc}
\usepackage{xcolor}
\usepackage{lineno}
\usepackage{hyperref}

\usepackage[letterpaper,top=2cm,bottom=2cm,left=2cm,right=2cm,marginparwidth=1.75cm]{geometry}

\usepackage{graphicx}
\usepackage{amsmath,amssymb,bm}
\usepackage{parskip}
\usepackage{multirow}

\journal{ }
\begin{document}

\title{Mesh-based Super-Resolution of Fluid Flows with Multiscale Graph Neural Networks}

\author[label1]{Shivam Barwey\texorpdfstring{\corref{cor1}}}
\author[label1]{Pinaki Pal}
\author[label2]{Saumil Patel}
\author[label3]{Riccardo Balin}
\author[label3]{Bethany Lusch}
\author[label3]{Venkatram Vishwanath}
\author[label4,label5]{Romit Maulik}
\author[label2]{Ramesh Balakrishnan}

\affiliation[label1]{organization={Transportation and Power Systems Division, Argonne National Laboratory},
            city={Lemont},
            postcode={60439},
            state={IL},
            country={USA}}

\affiliation[label2]{organization={Computational Science Division, Argonne National Laboratory},
            city={Lemont},
            postcode={60439},
            state={IL},
            country={USA}}
            
\affiliation[label3]{organization={Argonne Leadership Computing Facility, Argonne National Laboratory},
            city={Lemont},
            postcode={60439},
            state={IL},
            country={USA}}

\affiliation[label4]{organization={Mathematics and Computer Science Division, Argonne National Laboratory},
            city={Lemont},
            postcode={60439},
            state={IL},
            country={USA}}

\affiliation[label5]{organization={College of Information Sciences and Technology, Pennsylvania State University},
            city={University Park},
            postcode={16802},
            state={PA},
            country={USA}}

\cortext[cor1]{Corresponding author. E-mail address: {sbarwey@anl.gov} (S. Barwey).}

\begin{abstract}
A graph neural network (GNN) approach is introduced in this work which enables mesh-based three-dimensional super-resolution of fluid flows. In this framework, the GNN is designed to operate not on the full mesh-based field at once, but on localized meshes of elements (or cells) directly. To facilitate mesh-based GNN representations in a manner similar to spectral (or finite) element discretizations, a baseline GNN layer (termed a message passing layer, which updates local node properties) is modified to account for synchronization of coincident graph nodes, rendering compatibility with commonly used element-based mesh connectivities. The architecture is multiscale in nature, and is comprised of a combination of coarse-scale and fine-scale message passing layer sequences (termed processors) separated by a graph unpooling layer. The coarse-scale processor embeds a query element (alongside a set number of neighboring coarse elements) into a single latent graph representation using coarse-scale synchronized message passing over the element neighborhood, and the fine-scale processor leverages additional message passing operations on this latent graph to correct for interpolation errors. Demonstration studies are performed using hexahedral mesh-based data from Taylor-Green Vortex {\color{black}and backward-facing step} flow simulations at Reynolds numbers of 1600 and 3200. Through analysis of both global and local errors, the results ultimately show how the GNN is able to produce accurate super-resolved fields compared to targets in both coarse-scale and multiscale model configurations. Reconstruction errors for fixed architectures were found to increase in proportion to the Reynolds number. {\color{black}Geometry extrapolation studies on a separate cavity flow configuration show promising cross-mesh capabilities of the super-resolution strategy.}

\begin{keyword}
Graph neural networks \sep Super-resolution \sep Fluid dynamics \sep Taylor-Green Vortex \sep {\color{black}Backward-facing step} \sep Deep learning
\end{keyword}

\end{abstract}

\maketitle
\tableofcontents

\section{Introduction}
\label{sec:introduction}

Numerical simulations of unsteady fluid flows are critical to many engineering applications. However, to be truly predictive, they must resolve all relevant length and time scales in the governing partial differential equations (PDEs) - specifically, the Navier-Stokes equations in the context of fluid dynamics. Consequently, for applications characterized by quantities of interest influenced by both large and small-scale interactions, such high-fidelity simulations become prohibitively expensive when reliable and trustworthy numerical predictions are required \cite{moin1998direct}. As a result, there has been significant research in the past several decades towards developing models for accelerating multi-physics fluid dynamics simulations, particularly for turbulent flows \cite{zhiyin2015large,pitsch_arfm}. A popular strategy is large-eddy simulation (LES), which eliminates prohibitive spatiotemporal scales by filtering the governing equations. The filtered PDEs are solved on a significantly coarsened grid, which in-turn leads to orders-of-magnitude increases in allowable time-step sizes relative to fully-resolved simulations. The reduction framework provided by LES allows for a wide variety of \textit{coarse-grid} closure models predicting the effects of unresolved terms on the coarse-grid dynamics without ever explicitly constructing quantities on a resolved/fine grid. Conventional physics-based models in this area rely on phenomenological algebraic relations for such closures \cite{germano_smag,park2006dynamic}; enhancing (or even replacing) these models with data-driven corrections, through a combination of machine learning tools and automatic differentiation, is now an active and promising area of research \cite{kochkov2021machine,romit_differentiable,kolter_diffgnn}.

An alternative perspective (and related to the scope of this work) is one of super-resolution \cite{taira_sr_2023}. Here, the goal is to explicitly produce high-resolution flow-fields on a fine/resolved grid directly using deterministic model forms or statistical upsampling operators that bypass expensive direct numerical simulations. In fluid dynamics, the super-resolution strategy is useful from the following perspectives: (1) turbulence closure modeling (particularly from the angle of producing optimal LES closures \cite{langford_moser_jfm}), (2) providing initialization methods in adaptive mesh refinement simulations, (3) flow visualization and compression, and (4) investigating fundamental interactions between turbulence and other physical processes (e.g., shocks, flow separation) when simulating turbulence directly is too expensive. Super-resolution strategies in fluid dynamics have evolved over several decades, incorporating both purely physics-based methods and data-driven techniques. Before delving into the specific contributions of this work, the following text provides a brief overview of the super-resolution literature relevant to fluid dynamics modeling. For more comprehensive reviews, readers are directed to Refs.~\cite{taira_sr_2023,liu2020deep}.

A token example of physics-based super-resolution, inspired in part by techniques used in the computer vision and image deblurring fields, is the approximate deconvolution (AD) strategy \cite{stolz_ad}. Through scale similarity assumptions similar to those invoked in dynamic LES models \cite{germano_smag}, the AD process recovers fine-scale flow-fields using iterative application of a linear filtering operator on residual velocity fields; this is accomplished by approximating an inverse filtering operation on the fine grid. This baseline iterative approach (known as the Van Cittert algorithm) has since been improved using more robust numerical schemes in recent years \cite{san_vedula_deconv}. Alternative physics-based strategies leveraging inertial manifold assumptions \cite{titi_aim,maryam_jcp} have also been used for turbulent flow super-resolution; here, the unresolved scales are recovered through an instantaneous dynamics assumption (a manifold) in a spectral space upon projection of the governing equations onto eigenvectors of the diffusion operator. Variational multiscale methods present another class of physics-based strategies to recover unresolved quantities in a similar way, but can instead leverage localized projection operations to decompose the equations into coarse-scale and fine-scale contributions, after which the fine-scale models are recovered \cite{hughes_2007}. Statistical super-resolution strategies that are physics-based include the digital filtering approach \cite{klein2003digital} and Fourier-based turbulence generation \cite{karweit1991}. These methods allow one to sample the distribution of fine-scale flow-fields constrained to a target turbulence quantity of interest (such as integral length scale, turbulent kinetic energy, or the energy spectrum), and are used extensively to model turbulent boundary conditions and flow stratification \cite{wu_arfm}.

In recent years, data-driven super-resolution methods have gained considerable traction due to the increasing availability of high-fidelity simulation and real-world datasets. The inherent advantage of these approaches is that the super-resolution model parameters can be optimized to satisfy high-fidelity data statistics and flow properties directly, allowing such models to achieve promising results in non-canonical flow settings where many purely physics-based strategies break down \cite{taira_sr_2023}. Early applications showcased the promising potential of neural networks to address limitations associated with filtering assumptions in approximate deconvolution procedures -- for example, Refs.~\cite{maulik2017neural,romit_deconv_pof} develop a data-driven super-resolution strategy for turbulent flows using localized multi-layer perceptrons for this purpose. Since then, inspired by the success of convolutional neural networks (CNNs) in analogous computer vision tasks \cite{dong2015image}, the application of CNNs in the super-resolution of fluid flows has been successful in a variety of experimental \cite{wang2022deep,guo2022super,zhou2024large} and numerical \cite{gao_2021,fukami2019super,matsuo2021supervised} fluid dynamics settings. Extensions to not only spatial, but spatiotemporal upsampling, is also possible in this framework \cite{fukami2021machine,ren2023physr}. Transformer-based super-resolution models are also gaining traction in the fluid dynamics community \cite{xu2023super,wang2022transflownet}, although their overall performance advantages over conventional CNN-based architectures for these applications remain unclear.  

It should also be highlighted that generative modeling is particularly useful for statistical super-resolution, in that it allows one to achieve the modeling goal from the perspective of sampling from the distribution of high-resolution target fields conditioned on low-resolution or coarse fields (i.e., such models become high-dimensional conditional distribution sampling tools). For example, Ref.~\cite{malik_gan} applies CNN-based generative adversarial networks to achieve such probabilistic super-resolution in a statistically consistent manner, ensuring that the distribution of sampled high-fidelity fields adhere to the conditional statistics based on the low-fidelity input. More recent work in probabilistic super-resolution applies similar concepts, but with diffusion models as the backbone \cite{shu2023physics,shan2024pird,lienen2023generative}, mirroring in part the recent success and exposure of this strategy in the broader field of generative artificial intelligence.

Although promising in the context of extending capabilities of purely physics-based super-resolution models, many of these strategies encounter challenges when applied to complex geometries. Engineering applications often involve simulation configurations described by intricate geometries, where fluid flow data is represented on unstructured grids. As a result, approaches that depend on artificial or convolutional neural network architectures, and even modern vision transformers \cite{han2022survey}, are inherently incompatible with such unstructured representations. This limitation has spurred the development of \textit{mesh-based} data-driven models grounded in geometric deep learning \cite{gdl}. At the core of these models is the graph neural network (GNN), which consists of a series of message passing layers \cite{gilmer}. These layers take as input a graph connectivity (or adjacency) matrix, along with a mesh-based representation of the flow data. By leveraging this connectivity, GNNs can effectively capture nonlinear and non-local interactions between nodes and their respective neighborhoods. Additionally, this framework is naturally compatible with complex geometries, as unstructured flow-fields and non-uniformly arranged point clouds can be easily represented as graphs. Owing to this natural compatibility and the similarity between message passing layers and traditional numerical integration schemes used in computational fluid dynamics (CFD), GNN-based models have gained significant traction in the fluid dynamics community in recent years \cite{meshgraphnet,farimani_gnn,kolter_diffgnn}. The success of these initial GNN studies has since led to the development of enhanced multiscale message passing architectures that more efficiently model neighborhood relationships across larger length scales, improving single-scale GNN predictions \cite{multiscale_mgn,lino_2021,shivam_jcp,deshpande2024}. 

Development of GNN-based models in the fluid dynamics community has largely centered around surrogate forecasting \cite{meshgraphnet,graphcast,lino_2021,varun_gnn}, sub-grid modeling \cite{kolter_diffgnn,romit_differentiable}, and unstructured autoencoding applications \cite{shivam_jcp,hesthaven_jcp}. Little focus has been placed on leveraging the capabilities of GNN-based architectures for super-resolution of fluid flows, especially in three spatial dimensions. An exception can be found in Ref.~\cite{he2022flow}, which applies GNNs for reconstruction in two spatial dimensions; however, this work focuses on reconstruction of flow-fields from a set of sensor measurements, and does not explicitly move from coarse to fine grid representations. The goal of the present work is to fill this gap and develop a novel multiscale GNN architecture for the mesh-based super-resolution task applied to arbitrary three-dimensional (3D) mesh-based discretizations. The modeling framework, inspired by the recent physics-informed super-resolution work of Ref.~\cite{karthik_vms_superres}, operates within a localized patch-based approach and is conceptually in-line with p-refinement strategies leveraged in finite element simulations \cite{zienkiewicz1989effective}. 

{\color{black} 
In mesh-based GNN approaches, the initial step is to convert the underlying mesh into a graph consisting of nodes and edges. Considering the connection between local connectivity and numerical stencils, there are many ways to construct mesh-based graphs. These range from point-cloud-based graphs \cite{meshgraphnet,lino_2021} to finite volume-based \cite{karthik_gnn,shivam_jcp,li2025learning} and finite element-based graphs \cite{jaiman_fem_gnn}, among others. The GNN methodology introduced here centers on (a) constructing graphs inspired by spectral-element-discretized meshes at different element discretization levels, and (b) performing super-resolution through multiscale message-passing layers localized to coarse spectral-element neighborhoods on these graphs. Broadly, in the current work, message passing on a coarse spectral element (and its neighborhood) corresponds to information exchange along the edges connecting mesh vertices. On fine discretizations, message passing instead occurs within the element interior and across its faces via information propagation among internal quadrature points. Element locality is implicitly integrated into the model by limiting the operational scope during training to a single query element; that is, given a coarse query element (and, optionally, its neighborhood of mesh elements), the model is trained to predict some quantity of interest (in this case, a super-resolved field) within that same element.

It should be noted that the element-local model form used here is similar to the recently introduced $\phi$-GNN framework in Ref.~\cite{jaiman_fem_gnn}. That framework constructs finite-element-inspired hypergraphs and incorporates an element-local update step in the message passing layer designed to mimic the effect of local stiffness matrix calculations. Although the two approaches differ in graph interpretation and message passing formulations, both rely on the core idea of leveraging finite/spectral element-inspired graphs to impose physical locality into the model as an inductive bias and/or physical constraint. Although not explored in this work, combining local stiffness-matrix-inspired message passing layers \cite{jaiman_fem_gnn} with the multiscale strategy introduced here could further improve spectral accuracy in super-resolution.
}

{\color{black} The contributions of this work are as follows:
\begin{itemize}
    \item A new multiscale graph neural network architecture is introduced to handle mesh-based super-resolution. Specifically, coarse-scale and fine-scale message passing operations, separated by an intermediary unpooling layer, are used to recover the fine-scale flow-field given a coarse-scale flow-field as input. 

    \item {\color{black} Building on previous work \cite{karthik_vms_superres,fidkowski_sr,jaiman_fem_gnn}}, an element-local model form is adopted, which enables the construction of a training set using a relatively small number of snapshots. The element-local formulation used in this work is mesh-based, leveraging graph representations of mesh neighborhoods of spectral elements. 

    \item Using the GNN architecture, super-resolution analysis is performed in the following contexts: assessment of the effect of coarse element neighborhood size on the super-resolution accuracy, and characterization of the role of coarse and fine-scale message passing on fine-scale flow reconstruction. 

    \item To facilitate mesh-based graph representations in a manner similar to spectral element discretizations, the baseline message passing layer, popularized for fluid flows in Ref.~\cite{meshgraphnet}, is modified here to account for synchronization of coincident graph nodes, rendering compatibility with commonly used finite element-based mesh connectivities.
    
    \item Demonstrations are performed on complex mesh-based physical configurations in three spatial dimensions. Specifically, super-resolution capability is independently assessed for a Taylor-Green Vortex and Backward-Facing Step flow at Reynolds numbers of 1600 and 3200, and a separate Cavity flow configuration is used to demonstrate geometry extrapolation capability of trained models. 
\end{itemize}
}

The rest of the paper is organized as follows. In Sec.~\ref{sec:dataset}, the flow configuration, flow solver, mesh, and dataset used in this study are described. In Sec.~\ref{sec:methodology}, methodological details on the modeling goal, graph generation procedure, and GNN architecture are provided. Results are then detailed in Sec.~\ref{sec:results}, followed by concluding remarks in Sec.~\ref{sec:conclusion}

{\color{black} 
\section{Description of Datasets}
\label{sec:dataset}
Training and evaluation datasets in this work are generated from high-fidelity computational fluid dynamics (CFD) simulations in three different configurations: (1) the Taylor-Green Vortex (TGV), (2) flow over a backward-facing step (BFS), and (3) flow over a cavity. Before detailing the datasets for each configuration, the governing equations, simulation procedure, and solution representation are first described.

\subsection{Simulation Procedure}
For all three configurations, the governing equations are the incompressible Navier-Stokes (NS) equations in three spatial dimensions (with isothermal and Newtonian assumptions and no body force), given by
\begin{equation}
    \frac{\partial {\bf u}}{\partial t} + {\bf u} \cdot \nabla {\bf u} = -\nabla p + \frac{1}{\text{Re}}\nabla^2 {\bf u},
\end{equation}
\begin{equation}
    \nabla \cdot {\bf u} = 0.
\end{equation}
In the above equations, ${\bf u} = {\bf u}({\bf x},t)$ represents a three-dimensional fluid velocity defined at a particular point in physical space ${\bf x}$ and time $t$, the quantity $p = p({\bf x},t)$ is the scalar pressure, and $\text{Re}$ denotes the Reynolds number, a critical parameter that sets both the turbulence intensity and degree of scale separation observed in the fluid flow. To observe complex turbulent flow with appreciable scale separation in the above mentioned configurations, one must set $\text{Re}$ to appreciably high values. As such, in this work, two values for Reynolds number are considered to generate training and evaluation datasets: $\text{Re}=1600$ and $\text{Re}=3200$. In the case of the BFS and cavity, these Reynolds numbers are defined with respect to the step and cavity heights, respectively.

Numerical simulations are performed here with NekRS \cite{nekrs}, an extensively verified open-source GPU-based exascale CFD code developed at Argonne National Laboratory. NekRS solves the governing equations using a spectral element discretization in space on unstructured meshes, which can be composed of elements consisting of wedges, tetrahedra, and hexahedra (a regular mesh composed of hexahedral elements is used in this work) \cite{fischer_book}. Interested readers are directed to Ref.~\cite{nekrs} for further details of solver numerics implementation, time-stepping schemes, and scalability. 

At a particular Reynolds number, the flow solver generates a high-dimensional simulation trajectory describing the time-evolution of a velocity field on a set of spatial discretization points. These discretization points are prescribed by a mesh. Similar to conventional mesh-based simulation strategies (e.g., as in finite volume and finite element discretizations), in the spectral element method formulation of NekRS, the mesh is represented by a series of non-overlapping elements. In the solution procedure, the velocity field is solved on a set of quadrature points within (and at the interface of) each element. Examples of individual elements extracted from the TGV mesh (the same also applies to BFS and cavity meshes) are shown in Fig.~\ref{fig:ic_mesh_element}(c) -- the figure illustrates how, for the same underlying mesh, different levels of fidelity can be achieved through prescription of a different number of quadrature points within each element. NekRS leverages a Guass-Lobatto-Legendre (GLL) quadrature \cite{nekrs}, resulting in non-uniformly arranged spatial points within each element for high-order discretizations. The fidelity of the quadrature is characterized by the polynomial order, $P$, resulting in $N_p = (P+1)^3$ points per element in 3D. Shown in Fig.~\ref{fig:ic_mesh_element}(c) are two such polynomial order settings: (1) $P=1$, which corresponds to the \textit{coarse} flow-fields resulting in 8 GLL points per element, all of which are coincident with the mesh vertices, and (2) $P=7$, which represents \textit{fine} flow-fields resulting in 512 GLL points per element. 

An instantaneous mesh-based flow-field consistent with the notion of element discretization can be formally given by
\begin{equation}
\label{eq:flowfield}
    \mathbf{Y}(t) = 
    \left [ \,
        \begin{array}{c}
        {\bf y}_1(t) \\
        \hline
        {\bf y}_2(t) \\
        \hline
        \vdots \\
        \hline
        {\bf y}_{N_e}(t)\\
        \end{array}
    \, \right] \in \mathbb{R}^{(N_e N_p) \times 3}, 
    \quad {\bf y}_i(t) = 
    \left[
    \begin{array}{c}
    {\bf u}_i({\bf x}_1,t) \\
    \vdots \\
    {\bf u}_i({\bf x}_{N_p},t) \\
    \end{array}
    \right] \in \mathbb{R}^{N_p \times 3}, 
    \quad i = 1,\ldots,N_e. 
\end{equation}

\begin{figure}
    \centering
    \includegraphics[width=\columnwidth]{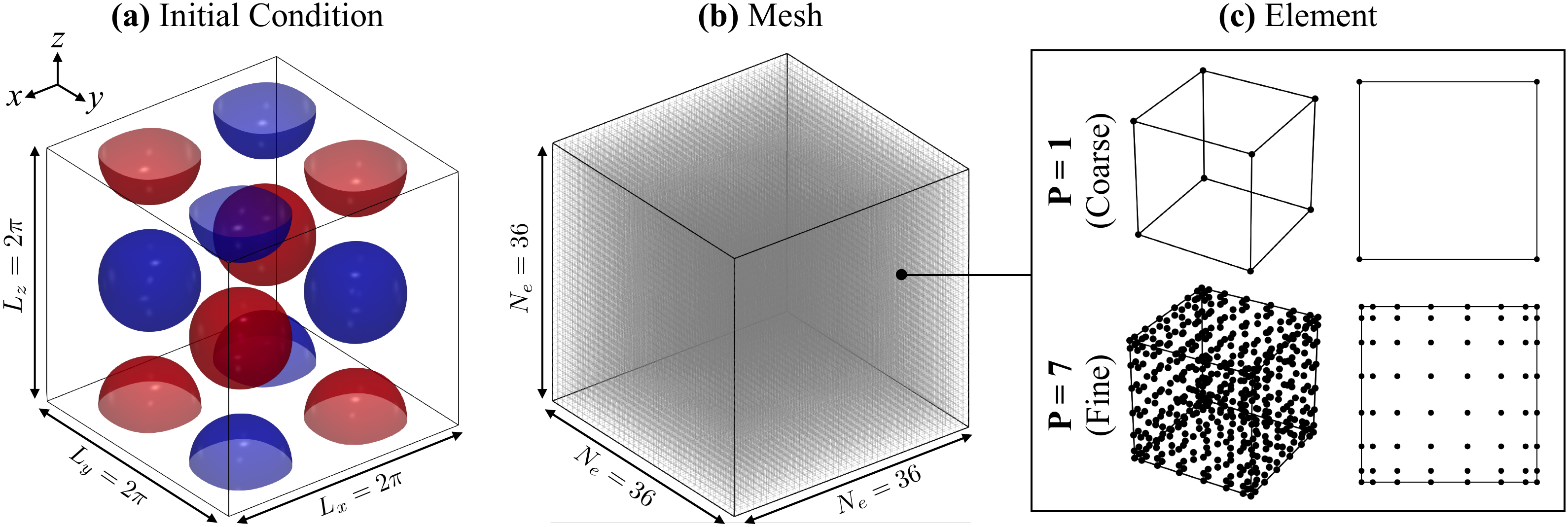}
    \caption{\textbf{(a)} Visualization of the TGV initial condition using contours of z-component of vorticity ($w_z$), with $w_z=1$ in red and $w_z=-1$ in blue. \textbf{(b)} Computational mesh ($36^3$ elements) used in NekRS simulations. \textbf{(c)} Visualization of GLL nodes in a single coarse (P=1, top) and fine (P=7, bottom) element.}
    \label{fig:ic_mesh_element}
\end{figure}

In Eq.~\ref{eq:flowfield}, ${\bf Y}(t)$ is the full flow-field at time $t$, consisting of a total of $N_e N_p$ discretization points, where $N_e$ is the total number of mesh elements and $N_p$ is the total number of quadrature points per element. Per Eq.~\ref{eq:flowfield}, each flow-field ${\bf Y}(t)$ is composed of a concatenation of element-local flow-fields, where ${\bf y}_i(t)$ denotes the velocity field in the $i$-th element. Distinctions between flow-field representations at different polynomial orders on the same mesh are indicated by subscripts where appropriate: a P=7 flow-field is denoted ${\bf Y}_7(t)$ (with the element-local quantity given by ${\bf y}_{7,i}(t)$). It should be noted that the flow-field representation in Eq.~\ref{eq:flowfield}, while consistent with the spectral element discretization employed in NekRS, also can be readily generalized to other mesh-based discretizations such as those found in finite-volume and discontinuous Galerkin methods (the finite-volume case, for example, results in $N_p = 1$). 

In all configurations described below, to facilitate a supervised training strategy (to be described in detail in Sec.~\ref{sec:methodology}), a coarse-fine flow-field pairing at a particular time instant $t$ is generated through a coarsening operation on a set of target P=7 flow-fields, producing a coarsened counterpart on the P=1 mesh (referred to as the projected target solution). This coarsening operation is expressed by a linear restriction operator ${\cal R}$, such that ${\bf Y}_1(t) = {\cal R} {\bf Y}_7(t)$. When moving from $P>1$ to $P=1$, the operator $\cal R$ in the spectral element framework coincides with an indexing operation on the flow-fields, since the $P=1$ element GLL points are a subset of the $P>1$ GLL points. However, without loss of generality, $\cal R$ can be interpreted as a general non-invertible filter acting on an instantaneous flow-field \cite{langford_moser_jfm}. 

\subsection{Taylor-Green Vortex (TGV)}
\label{sec:tgv_description}
The TGV is a classic benchmark problem in the CFD literature, with early numerical studies performed by Orszag \cite{orszag1974numerical} and Brachet et al \cite{brachet1983small}. It has since been used in a wide variety of applications related to the numerical simulation of inviscid and turbulent fluid flows; recent use-cases include, but are not limited to: (a) spurring the development of low-dissipation numerical schemes for flow solvers in complex geometries \cite{pirozzoli_openfoam}, (b) usage as a benchmark configuration to evaluate CFD solver numerics \cite{johnsen_jcp,diosady2015case,thevenin_cf}, and (c) facilitating the development of super-resolution and turbulence models \cite{san_vedula_deconv,bao2022physics,klein_tgv}, as done here.

The TGV initial condition is shown in Fig.~\ref{fig:ic_mesh_element}(a), and a schematic of the mesh used for the TGV simulation is shown in Fig.~\ref{fig:ic_mesh_element}(b). The total number of mesh elements is fixed to $36^3$, resulting in roughly 24M discretization points on the fine P=7 mesh\footnote{An element polynomial order of $P=7$ on this mesh is roughly equivalent to a conventional $256^3$ structured grid discretization, which is deemed to satisfy the requirements for fully resolved TGV simulations at the two Reynolds numbers considered here \cite{brachet1983small}.} and 0.37M points on the coarse P=1 mesh. A description of the physical characteristics of the TGV is provided in \ref{sec:app:tgv} for the unfamiliar reader. Ultimately, the datasets used in the training and evaluation procedure are sourced from temporal ranges characterized by turbulent flow behavior near the peak dissipation rate for both Re=1600 and 3200.

Visualization of a fine-to-coarse flow-field pair is shown in Fig.~\ref{fig:coarsening_example}(a) and (b) for P=7 and P=1, respectively, for a $\text{Re}=1600$ snapshot. Consistent with expectations, the coarsening operation retains many of the large-scale flow characteristics (such as flow symmetry about the x-y quadrants); however, the elimination of fine-scale flow-features is evident, posing a challenging super-resolution task. This elimination is highlighted in the corresponding energy spectrum curves in Fig.~\ref{fig:coarsening_example}(c). The spectra explicitly highlight the elimination of high-wavenumber energy content through the coarsening procedure for both Reynolds numbers, as well as increase in high-energy content in the target flow-fields with Reynolds number. Training data for the TGV consists of three snapshots at $t = [8, 9, 10]$ time units, resulting in a total of $139968$ coarse-fine element pairs in the training set for each Reynolds number. 

\begin{figure}
    \centering
    \includegraphics[width=\columnwidth]{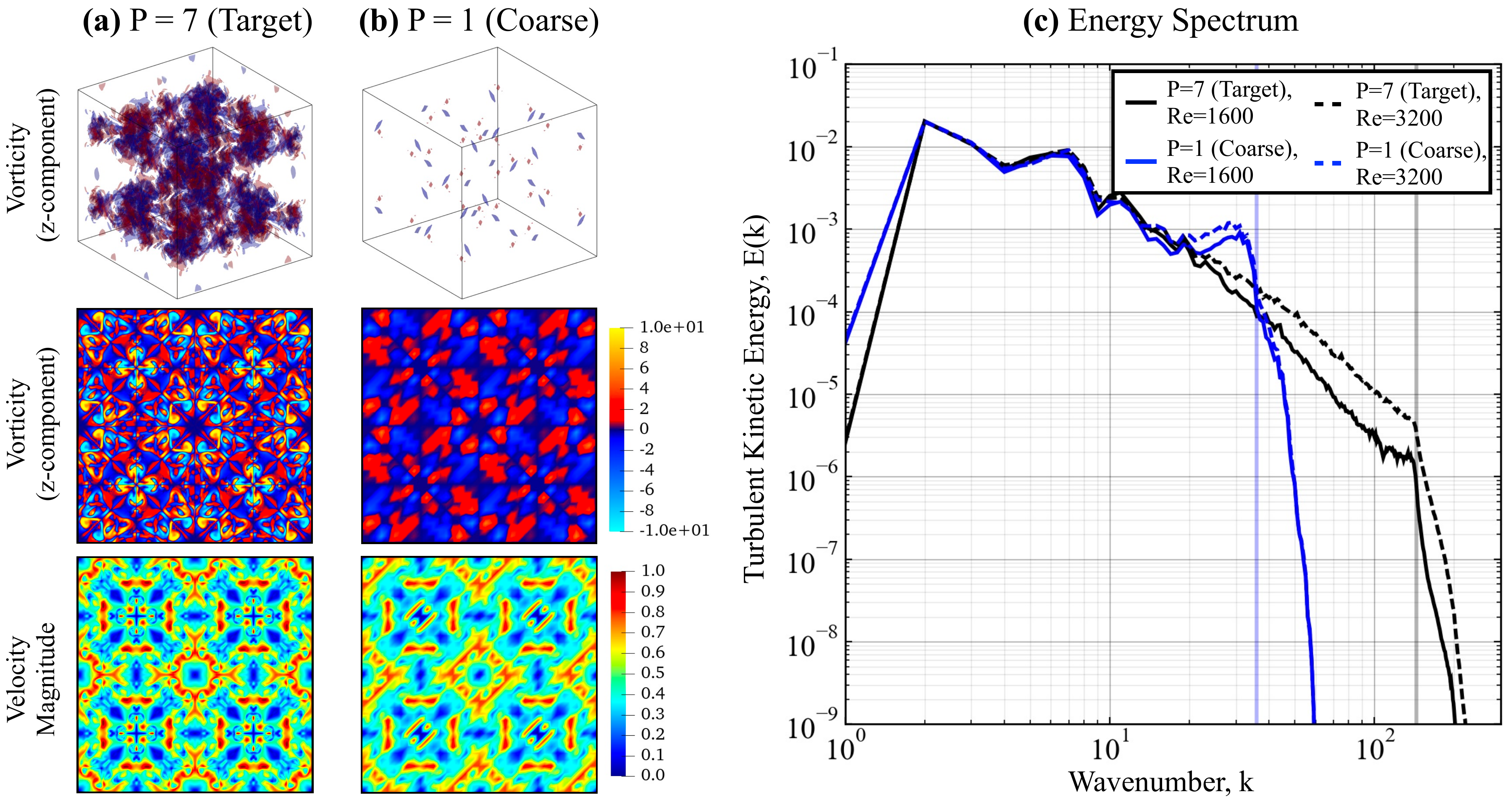}
    \caption{{\color{black}\textbf{(a)} Visualization of Re=1600 flow-field ${\bf Y}_7$ at $t=9$. Top plot shows vorticity contours, middle shows vorticity in the 2D x-y plane at z=$\pi/2$, and bottom shows velocity magnitude in the same plane. \textbf{(b)} Same as (a), but for coarsened flow-field ${\bf Y}_1$. \textbf{(c)} Turbulent kinetic energy versus wavenumber (energy spectrum) for ${\bf Y}_7$ (DNS, black) and ${\bf Y}_1$ (projected DNS, blue) at $t=9$. Solid and dashed curves correspond to Re=1600 and Re=3200, respectively.}}
    \label{fig:coarsening_example}
\end{figure}

\subsection{Backward-Facing Step (BFS) and Cavity}
\label{sec:bfs_cavity_description}

The BFS and cavity mesh configurations are shown in Fig.~\ref{fig:bfs_cavity_mesh}. These geometries have been studied in a number of experimental \cite{armaly1983experimental,kostas2002particle,ben1998cavity}, numerical \cite{lawson2011review,le1997direct}, and machine learning \cite{shivam_cmame,wang2023graph} works in the context of physical analysis and modeling of geometry-induced separated turbulent flows. In both cases, flow enters from the inflow (detailed in white arrows in Fig.~\ref{fig:bfs_cavity_mesh}) into a channel upon encountering a backward-facing step. In both the BFS and cavity, the step anchor triggers flow separation, resulting in a shear layer that then reattaches to the wall further downstream (in the case of the cavity, reattachment may also occur on the opposing vertical wall). The separation triggers a fully turbulent flow, with complex multiscale flow features downstream of the separation point, until the flow exits at the domain outlet. At higher Reynolds numbers, alongside increased turbulent intensity, flow reattachment points and other large-scale dynamical quantities of interest (such as recirculation and shear layer dynamics) become noticeably more complex.

The boundary conditions for both cases include fixed inflow and stabilized outflow \cite{dong2014robust} settings, span-wise periodicity, and no-slip walls everywhere else, with the exception of an initial \textit{ramp-up region} that extends along the top and bottom walls for a streamwise distance of $5h$ from the inlet (shown in the 2D schematics in Fig.~\ref{fig:bfs_cavity_mesh}). In this ramp-up region, slip wall boundary conditions are used (instead of no-slip) to mitigate discontinuity effects associated with a fixed-velocity inflow. Ultimately, in both BFS and cavity cases, the flow traverses a streamwise distance of $10h$ before encountering the step anchor -- the first half of this distance is the ramp-up phase, with no-slip conditions prescribed afterwards. 

\begin{figure}
    \centering
    \includegraphics[width=\linewidth]{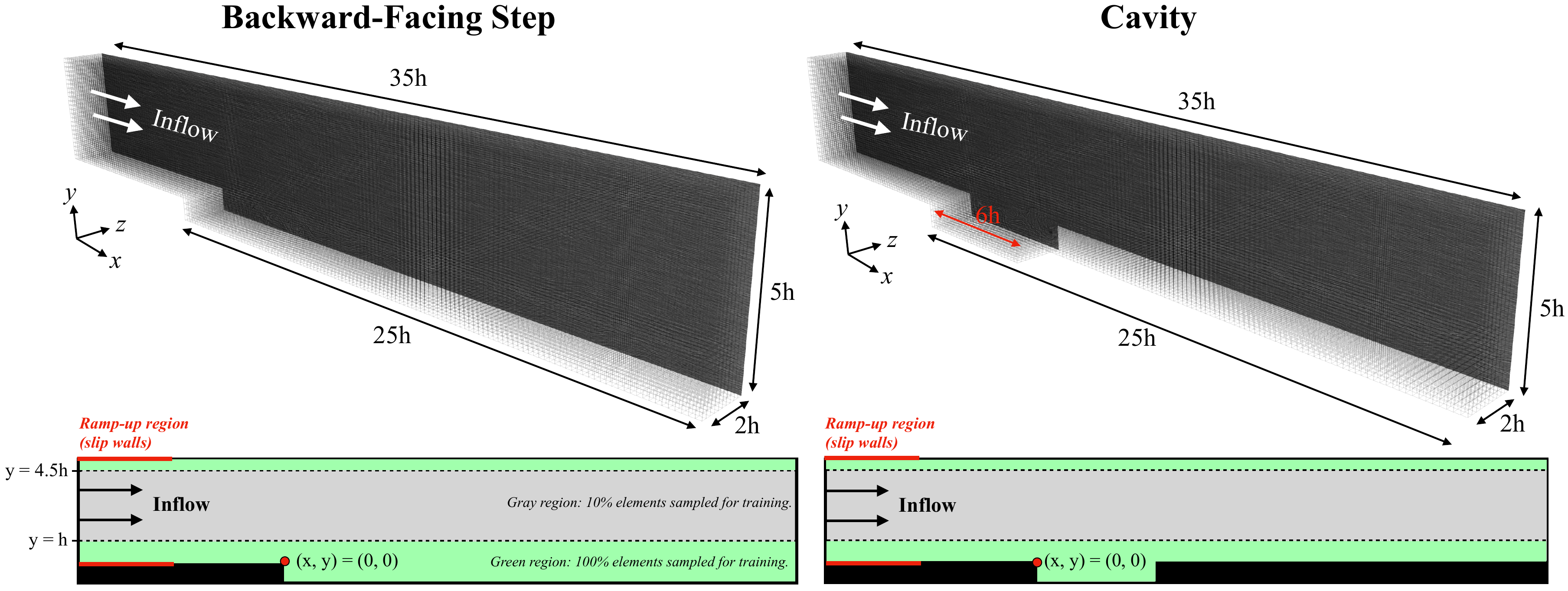}
    \caption{Meshes for backward-facing step (left) and cavity (right) configurations. Bottom schematics show 2D cross-sections.}
    \label{fig:bfs_cavity_mesh}
\end{figure}

The BFS and cavity meshes consist of a total of 46480 and 42440 hexahedral elements-per-snapshot, respectively. It should be noted that the P=7 target cases here are under-resolved relative to direct numerical simulation requirements (the resolution here is roughly equivalent to $h/28$, where $h$ is the step/cavity height). Despite this, for this study, the target resolution is adequate for capturing highly complex turbulent flow features in both configurations. 

This is confirmed through visualization of fine-coarse flow-field pairs, as is shown in Fig.~\ref{fig:bfs_cavity_coarse_fine}. Comparison of Fig.~\ref{fig:bfs_cavity_coarse_fine}(a) and (b) highlights the added complexity of increasing Reynolds number from 1600 to 3200 through the addition of smaller length-scales near and downstream of the reattachment point. Comparison of Fig.~\ref{fig:bfs_cavity_coarse_fine}(a) and (c) highlights the physical differences between BFS and cavity flow-fields at the same Reynolds number; the primary difference is due to flow impingement on the rear wall of the cavity (a reverse-step), in which additional recirculation zones are present in the place of a clean reattachment point to the bottom wall. 
    
\begin{figure}
    \centering
    \includegraphics[width=\linewidth]{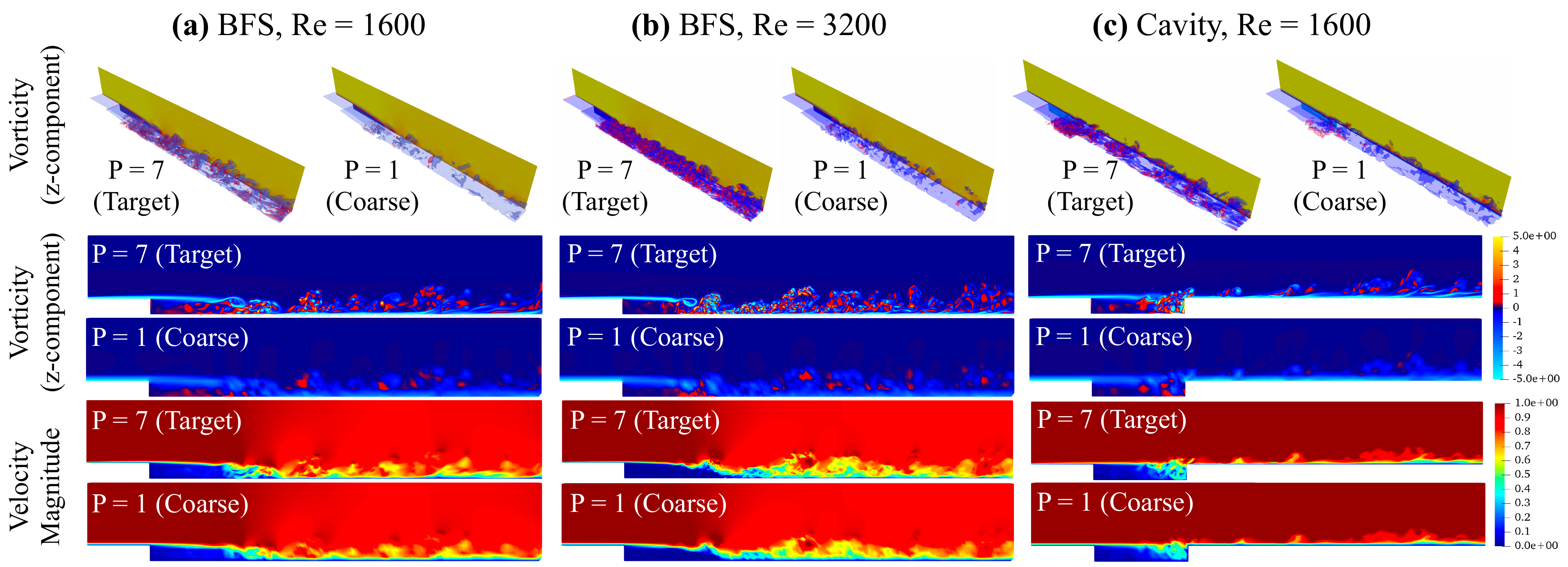}
    \caption{\textbf{(a)} Coarse-fine training snapshot pairs for BFS configuration at Re=1600, showing z-component vorticity contours (top), x-y plane slices of vorticity (z-component) taken at z=1 (middle), and x-y plane slices of velocity magnitude (bottom). \textbf{(b)} Same as (a), but for BFS at Re=3200. \textbf{(c)} Same as (a), but for cavity at Re=1600. Vorticity contours show $w_z=-5$ in red and $w_z=5$ in blue. Domains are cropped in x- and y-directions for near-step visualization.}
    \label{fig:bfs_cavity_coarse_fine}
\end{figure}

Training data for BFS and cavity configurations are sourced from a set of snapshots after $t=100$ time units of simulation, after which all of the initial vortex transients are shed. After this point, for each of the BFS and cavity cases, 10 training snapshots are extracted, each separated by $\Delta t = 2.5$ time units. From each of these snapshots, elements to populate the training dataset are sampled according to the following setup (illustrated in the lower schematics in Fig.~\ref{fig:bfs_cavity_mesh}): (a) 100\% of elements below $y=h$ (i.e., below one step-height above the step anchor) are retained, (b) 100\% of elements above $y=4.5$ are retained, and (c) 10\% of the remaining elements (largely in the freestream) are randomly sampled. This implicitly biases the training algorithm away from the freestream, and towards the boundary layer and turbulent flow features. 

After this procedure, the total number of training elements is 138060 for the BFS configuration and 103320 for the cavity, resulting in similar total training element counts for all three configurations. \textbf{Lastly, it is emphasized that the scope of the BFS and cavity configurations are constrained to the following}: (a) the BFS is leveraged as an additional demonstration case for GNN super-resolution performance, in that additional GNNs are trained and evaluated using data from this configuration; (b) the cavity is leveraged as a \textit{geometry extrapolation} testbed: both TGV- and BFS-trained models are evaluated on the cavity to determine geometry generalization capability of models trained from physically different data sources. 
}

\section{Methodology}
\label{sec:methodology}
With the flow configuration and dataset described in Sec.~\ref{sec:dataset}, this section proceeds by defining the super-resolution modeling goal, GNN architecture, and training setup.

\subsection{Modeling Scope}
\label{sec:modeling_scope}
In the deterministic sense, the super-resolution goal is to obtain an instantaneous mapping that lifts the low-resolution flow-field to its high-resolution counterpart. The conventional strategy is to achieve this modeling goal in a \textit{full-field} context, where the entire snapshot (all $N_e N_p$ spatial discretization points) is provided to the model in one go. In the context of the P=1 and P=7 element-based flow-fields described in Sec.~\ref{sec:dataset}, this produces a single input-target pair of $({\bf Y}_1(t), {\bf Y}_7(t))$ at time $t$.  

Instead of operating on a full-field representation, the super-resolution models in this work are constrained to an \textit{element-local} representation (also leveraged in Refs.~\cite{karthik_vms_superres,fidkowski_sr}), wherein a single input-target pair instead becomes $({\bf y}_{1,i}(t), {\bf y}_{7,i}(t))$, with ${\bf y}_{1,i}(t)$ (resp. ${\bf y}_{7,i}(t)$) denoting the instantaneous P=1 (resp. P=7) flow-field local to element $i$ at time $t$. Element-local approaches offer the following advantages: (1) imposition of an inductive bias in-line with physical expectations, in that reconstruction of the fine-scale information in a sub-domain of physical space (in this case, the element boundaries) is a function of coarse-scale information in and around the same region, and (2) direct compatibility with element-based meshes and discretizations. 

\begin{figure}
    \centering
    \includegraphics[width=\columnwidth]{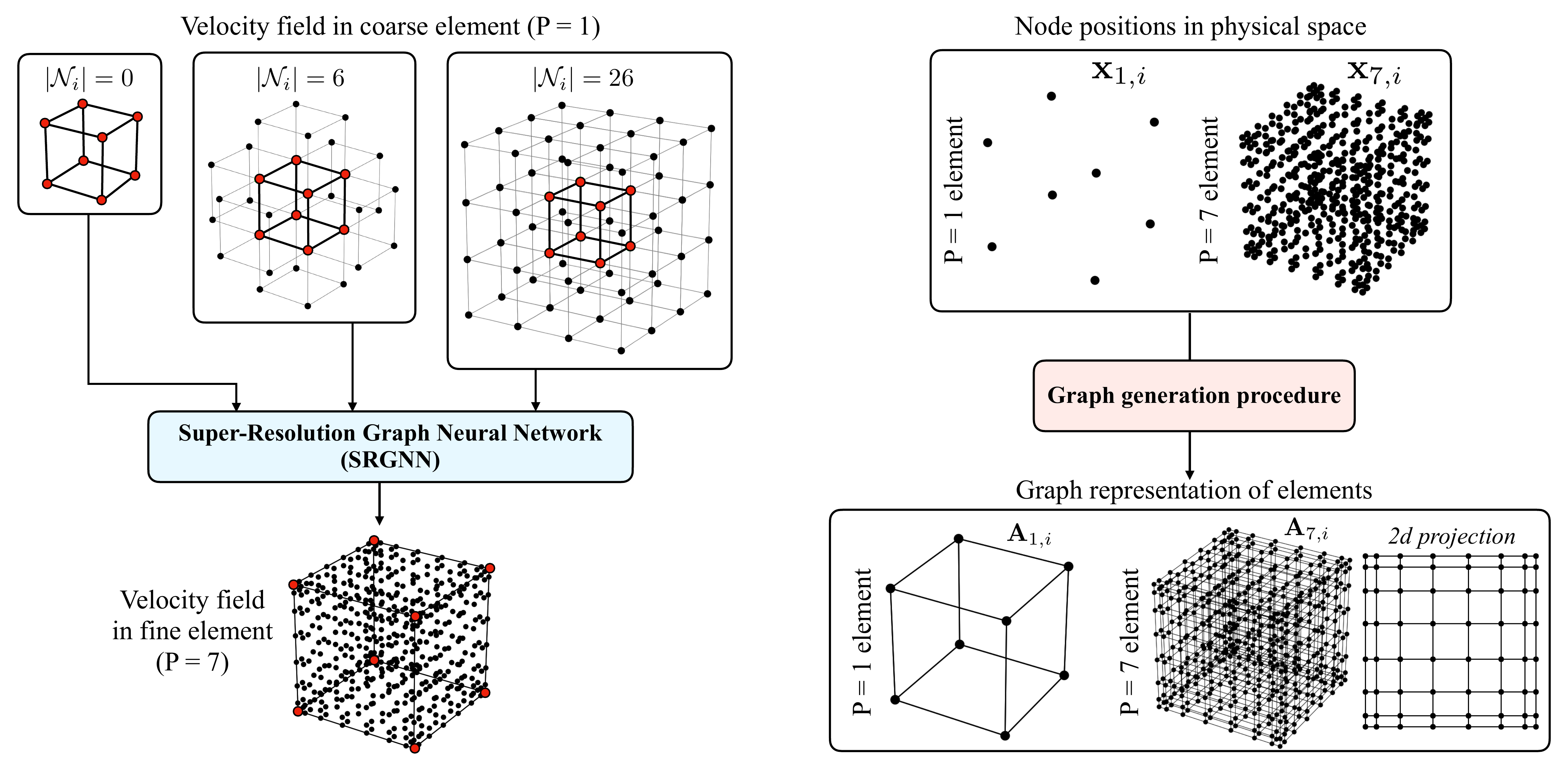}
    \caption{\textbf{(Left)} Illustration of the SRGNN modeling scope, as described in Sec.~\ref{sec:modeling_scope}. Shown are the three different sizes of coarse element neighborhoods considered here (0, 6, and 26), with central query element nodes marked in red. Coarse element neighborhood (whose graph node features contain velocity fields) is passed as input into the SRGNN, which outputs the super-resolved velocity field in the same query element. \textbf{(Right)} Illustration of element-local graph generation procedure for P=1 and P=7 discretizations, as described in Sec.~\ref{sec:gnn_architecture}.}
    \label{fig:scope_graphgen}
\end{figure}

As such, a super-resolution graph neural network (SRGNN) is introduced in the element-local framework, which allows for localized mesh-based recovery of high-resolution flow information from low-resolution inputs. For the $i$-th element, the SRGNN operation -- illustrated in Fig.~\ref{fig:scope_graphgen} -- is given by
\begin{equation}
    \label{eq:srgnn}
    \widetilde{\bf y}_{7,i} = \widehat{\bf y}_{7,i} + {\cal G}({\bf y}_{1,i}, {\cal N}_{i}, {\cal A}_i; \theta) = \widehat{\bf y}_{7,i} + \widetilde{\bf r}_{7,i},
\end{equation}
with time variable $t$ omitted for clarity. In Eq.~\ref{eq:srgnn}, $\cal G$ denotes a graph neural network described by the parameter set $\theta$. Through a training procedure, the modeling goal is to optimize the parameters in $\theta$ such that $\widetilde{\bf y}_{7,i} \approx {\bf y}_{7,i}$, where $\widetilde{\bf y}_{7,i}$ is the predicted high-resolution flow-field local to element $i$ and ${\bf y}_{7,i}$ is the corresponding target. The GNN $\cal G$ achieves this goal by modeling the residual with respect to the interpolated velocity field $\widehat{\bf y}_{7,i}$. In other words, $\widehat{\bf y}_{7,i} \leftarrow {\cal I}({\bf y}_{1,i})$, where $\cal I$ is a parameter-free interpolation/lifting function executed local to the element (details on this function -- which is taken here to be a linear interpolation based on a K-nearest neighbors (KNN) weighting -- are provided in Sec.~\ref{sec:unpool}). It should be noted that although a residual form is leveraged here, a direct model form can be recovered by omitting $\widehat{\bf y}_{7,i}$ in Eq.~\ref{eq:srgnn}. 

As shown in Eq.~\ref{eq:srgnn}, alongside the low-resolution velocities ${\bf y}_{1,i}$, the GNN takes two additional inputs, namely ${\cal N}_i$ and ${\cal A}_i$. The quantity ${\cal N}_i$ is a set containing a neighborhood of coarse element velocity fields conditioned on the position of the $i$-th input coarse element (referred to as the query (or central) element). The size of this neighborhood set is a hyperparameter; larger neighborhoods correspond to a higher level of coarse information content utilized in the super-resolution procedure. In this work, three neighborhood sizes are compared (illustrated in Fig.~\ref{fig:scope_graphgen}) to uncover the degree of neighboring-element influence on super-resolution accuracy: $|{\cal N}_i| = 0$ (no coarse element neighbors considered), $|{\cal N}_i| = 6$ (the neighborhood surrounds only the faces of the hexahedral query element), and $|{\cal N}_i| = 26$ (the neighborhood completely surrounds the query element). 

The quantity ${\cal A}_i = ({\bf A}_{1,i}, {\bf A}_{7,i})$ contains the element-local graph connectivity (or adjacency) matrices -- ${\bf A}_{1,i}$ is the connectivity between GLL quadrature points for the P=1 element and ${\bf A}_{7,i}$ for the P=7 element. These connectivities, produced in the graph generation procedure described in Sec.~\ref{sec:gnn_architecture}, prescribe the way in which information exchange occurs during GNN message passing evaluations. 

\subsection{Graph Generation} 
\label{sec:graph_gen}
Before presenting the GNN architecture, it is first important to describe the graph generation step required to convert the mesh-based flow-field into a graph representation. In broad terms, a graph is formally defined by the two-tuple $G=(V,E)$, where $V$ contains the set of nodes and $E$ the set of edges (the connectivity) prescribing connections between nodes. In the spirit of the element-local flow-field decomposition described above, the graph generation process is also conducted here in an element-local manner: for each snapshot composed of $N_e$ elements, the result of the graph generation process is a set of $G_i$ graphs, where $i = 1,\ldots,N_e$ represents the element index. 

Starting from a set of spatial discretization points (interpreted as a point cloud coinciding with graph vertices $V$), many strategies have been explored in the literature to populate the graph edges $E$ from mesh-based flow-fields. These strategies can be broadly categorized into two classes: (1) leveraging an edge generation algorithm, such as a nearest-neighbor method (which allows the user to fix the node degree) \cite{lino_2021} or a radius-based method (which allows the user to fix the edge length scale) \cite{meshgraphnet}, and (2) constructing a connectivity consistent with the numerical discretization procedure and underlying mesh. The latter pathway, which has been employed in recent work to produce graph connectivities inspired by finite volume \cite{karthik_gnn,shivam_jcp} and finite element \cite{jaiman_fem_gnn} discretizations, is leveraged here to produce connectivities in-line with the element-local discretizations used in the spectral element method.

More specifically, as illustrated in Fig.~\ref{fig:scope_graphgen} (right), the nodes for graph $G_i$ coincide with the spatial locations of the GLL quadrature points, and the undirected edges (used to populate the adjacency matrix ${\bf A}_{P,i}$ for an element of polynomial order $P$) are constructed to connect neighboring quadrature points based on a skewed and structured stencil within the element. In this setting, quadrature points (graph nodes) internal to the element have 6 neighbors, those on the element faces have 5 neighbors, and those at the element vertices have 3 neighbors. As shown in Fig.~\ref{fig:scope_graphgen}, in the P=1 case, this graph connectivity is equivalent to the hexahedral mesh itself; in the P=7 case, the graph is characterized by edges at notably smaller length scales. This disparity in edge length scales in coarse and fine element graphs forms the basis for the multiscale GNN strategy described below. Ultimately, constructing edges between the GLL quadrature points in this fashion leads to an additional inductive bias baked into the super-resolution model: within a single element, spatial discretization points nearby one-another in physical space influence each other more than those further away. 

\subsection{Graph Neural Network Architecture} 
\label{sec:gnn_architecture}
\begin{figure}
    \centering
    \includegraphics[width=\columnwidth]{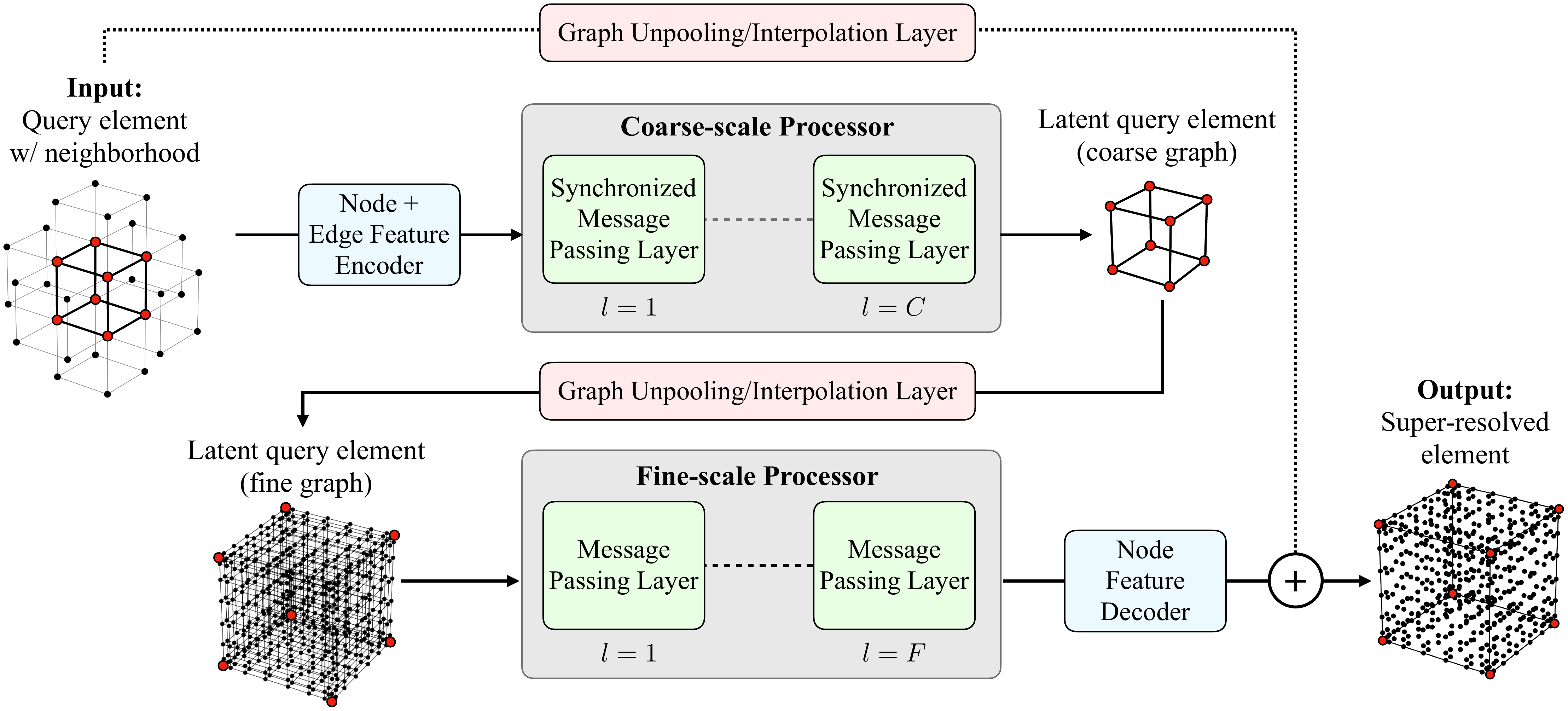}
    \caption{Super-resolution graph neural network architecture. Descriptions of each component provided in Sec.~\ref{sec:gnn_architecture}.}
    \label{fig:gnn_arch}
\end{figure}

The GNN architecture is illustrated in Fig.~\ref{fig:gnn_arch}, which depicts a flowchart for the forward pass operation used to generate the super-resolved flow-field within each element. The architecture, which falls within the class of encode-process-decode based GNN models \cite{meshgraphnet}, can be decomposed into the following four core components: (1) a node and edge-wise feature encoder, (2) a coarse-scale processor, (3) a graph unpooling layer, and (4) a fine-scale processor. These components are described in succession below.

To facilitate descriptions of these architecture components, notation for the graph node and edge features (as well as the edge feature initialization procedure) are first provided. In what follows, element indices are dropped for notational clarity -- all variables are assumed to exist in the context of a single element-based graph unless stated otherwise. A single node in an arbitrary element-based graph is described by the index $j$, where $j=1,\ldots,N_P$ (recall that $N_P = (P+1)^3$, where $P$ is the element polynomial order). The $j$-th node features consist of two three-component vectors: the velocity field and physical space position at the corresponding spatial location, denoted as ${\bf y}_j \in \mathbb{R}^3$ and ${\bf p}_j \in \mathbb{R}^3$, respectively. The edge features are given by ${\bf e}_{jk}$, where $j$ denotes the sender (or owner) node index and $k$ the receiver (or neighbor) node index. Edge features are initialized using relative velocities and positions (i.e., distances) as ${\bf e}_{jk} = ({\bf y}_j - {\bf y}_k, {\bf p}_j - {\bf p}_k, \lVert {\bf p}_j - {\bf p}_k \rVert _2^2)$. 

\subsubsection{Node/Edge Feature Encoder and Decoder} 
\label{sec:feature_encoder}
All initial node and edge features are encoded into a higher-dimensional latent space using independently parameterized multi-layer perceptrons (MLPs). This operation is denoted by the node-wise operation ${\bf y}_j \leftarrow \text{MLP}({\bf y}_j)$ and the edge-wise operation ${\bf e}_{jk} \leftarrow \text{MLP}({\bf e}_{jk})$, batched over all nodes and edges in the graph respectively. The result of this encoding stage is a set of latent node and edge features defined by a single hidden channel dimensionality $N_H$ (i.e., all node and edge features exist in $\mathbb{R}^{N_H}$ after the action of the respective MLPs). The hidden channel dimensionality is set to $128$ here; as demonstrated in previous studies \cite{meshgraphnet,lino_2021,shivam_jcp}, enriching node and edge feature spaces in this manner results in a more expressive architecture, allowing subsequent GNN layers to more accurately capture nonlinear relationships within graph node neighborhoods. Node and edge features remain in the $\mathbb{R}^{N_H}$-space until the final node decoding stage, which utilizes another MLP to bring back the node features into the velocity representation in $\mathbb{R}^{3}$-space. Edge features are discarded in the end, as the objective function is based on node velocity values on the fine graph.

\subsubsection{Coarse-Scale Processor}
\label{sec:coarse_scale_processor}
The coarse-scale processor (CSP), through nonlinear aggregation procedures provided by message passing layers, transforms the input query element graph into a \textit{latent graph}. More specifically, given a coarse query element graph $G^q_c$ (with associated connectivity, node, and edge features) and its neighborhood of element graphs ${\cal N}(G^q_c)$, the action of the coarse-scale processor can be described by $\widetilde{G}^q_c \leftarrow \text{CSP}(G^q_c, {\cal N}(G^q_c))$, where $\widetilde{G}^q_c$ is the latent query graph. This latent graph still exists on the coarse scales; $\widetilde{G}^q_c$ retains the same node spatial positions and connectivity matrix as the input query graph $G^q_c$, but has modified node and edge features that encode the interactions between the node features in the input graph and its neighborhood ${\cal N}(G^q_c)$ through message passing operations.

Message passing was recently introduced as a modeling framework in Ref.~\cite{gilmer} to classify different types of layers leveraged in GNNs and geometric deep learning frameworks. As such, various message passing schemes can be utilized. The variant used here comes from Refs.~\cite{meshgraphnet,battaglia2018relational}, where each layer consists of three steps: (1) an edge update step in which edge features are updated using corresponding node features at the respective edge, (2) an aggregation step that sends the sum of updated edge features computed in step 1 to the corresponding owner nodes, and (3) a node update step in which the node features are updated using the result of the edge aggregation in step 2. Parameters are introduced into the layer by leveraging MLPs in steps 1 and 3. Stacking many such layers results in an expressive operation that models complex non-local interactions in graph neighborhoods through iterative cycles of node and edge feature updates.

\begin{figure}
    \centering
    \includegraphics[width=\columnwidth]{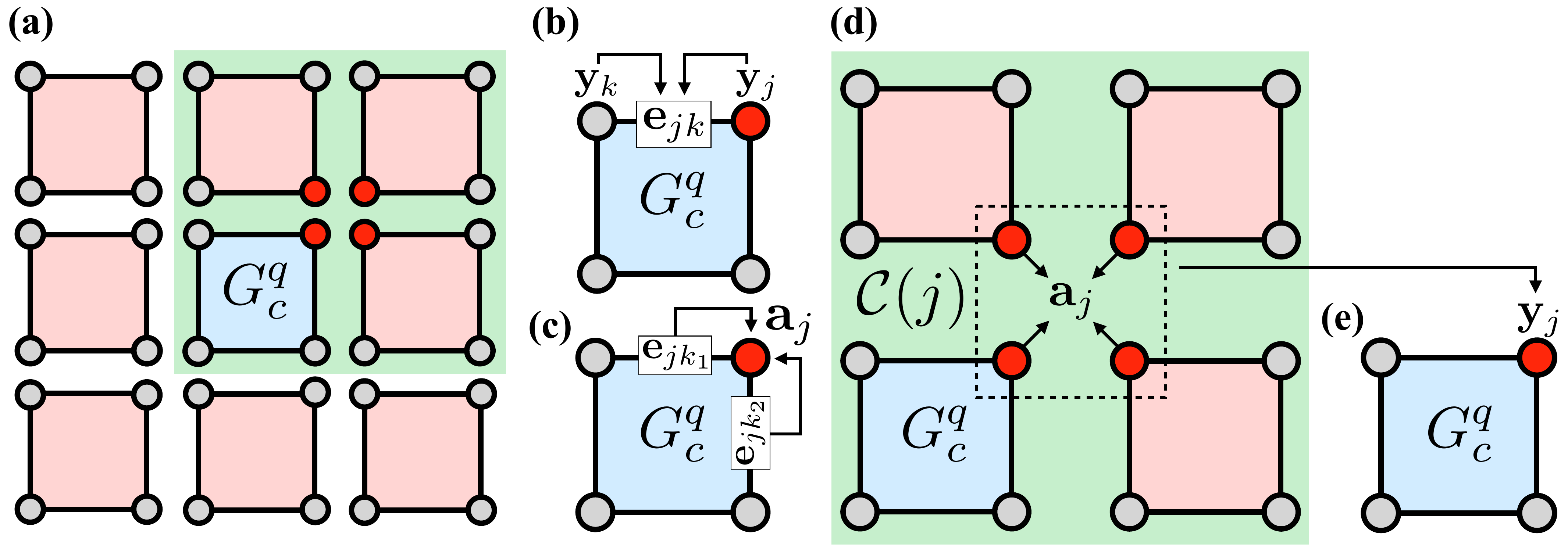}
    \caption{Schematic of synchronized message passing layer operations given by Eqs.~\ref{eq:mp_edge_update}-\ref{eq:mp_node_update}. For ease of visualization, 2D projections of P=1 elements are shown and level index superscripts are ignored. \textbf{(a)} Coarse query element $G_c^q$ (blue box) and its neighbors ${\cal N}(G_c^q)$ (red boxes) for the 26-neighbor case. Coincident owner nodes for purposes of illustration are denoted by red markers. \textbf{(b)} Edge update process (Eq.~\ref{eq:mp_edge_update}) on an edge in query element. \textbf{(c)} Edge aggregation process (Eq.~\ref{eq:mp_edge_agg}) for owner node in query element. \textbf{(d)} Synchronization of edge aggregates as described in Eq.~\ref{eq:mp_sync}, with ${\cal C}(j)$ indicated by dashed box. \textbf{(e)} Node update process (Eq.~\ref{eq:mp_node_update}) shown for owner node in query element. Note that in practice, the operations (b)-(d) are batched/parallelized over all nodes and edges on all graphs in the element neighborhood in (a).}
    \label{fig:sync_mp_layer}
\end{figure}

In the CSP, a set of $C$ message passing layers is executed on both the query graph $G^q_c$ and all graphs in ${\cal N}(G^q_c)$ in parallel. Note that since adjacency matrices are localized to each element in this framework (see Sec.~\ref{sec:graph_gen}), the three-step message passing procedure outlined above leads to an inconsistency when considering parallel evaluations in the element neighborhood, in that it does not allow for information exchange \textit{across} element-local graph boundaries. In other words, there are situations in which nodes in the query graph $G^q_c$ are coincident in physical space with nodes in neighboring element graphs -- physical consistency requires such coincident nodes to have identical node features throughout the message passing process. To enforce this constraint, a variation of the baseline message passing layer, referred to as a synchronized message passing layer, is introduced here. The synchronized message passing layer, which adds a new synchronization step between the previous steps 2 and 3 described above, is illustrated in Fig.~\ref{fig:sync_mp_layer} and is given by the following equations:
\begin{align}
    \text{Edge update:} &\quad 
    {\bf e}_{jk}^l = \text{MLP}_e^l ({\bf e}_{jk}^{l-1}, {\bf y}_j^{l-1}, {\bf y}_k^{l-1}),  \label{eq:mp_edge_update}\\
    \text{Edge aggregation:} &\quad 
    {\bf a}_{j}^l = \sum_{k \in {\cal M}(j)} {\bf e}_{jk}^l,  \label{eq:mp_edge_agg}\\
    \text{Synchronization:} &\quad 
    {\bf a}_{j}^l = \frac{1}{|{\cal C}(j)|} \sum_{k \in {\cal C}(j)}{\bf a}_k^l \label{eq:mp_sync}\\
    \text{Node update:} &\quad 
    {\bf y}_{j}^l = \text{MLP}_y^l ({\bf y}_j^{l-1}, {\bf a}_{j}^l). \label{eq:mp_node_update} 
\end{align}
The above equations operate in the context of element-local graphs, with subscripts $j$ and $k$ denoting graph node indices, and the superscript $l$ denoting the layer index. The functions $\text{MLP}_e^l$ and $\text{MLP}_y^l$ are the edge and node updater MLPs, respectively, at the $l$-th message passing layer. Equation~\ref{eq:mp_edge_update} provides the edge update operation, which explicitly depends on the edge features from the previous layer and the corresponding owner and neighbor node features at the given edge. The edge aggregation operation is shown in Eq.~\ref{eq:mp_edge_agg}, and leverages the updated edge features to produce an intermediary aggregate feature vector ${\bf a}_j^l$, which is a node-based quantity (i.e., Eq.~\ref{eq:mp_edge_agg} transforms edge-based features to node-based features through a reduction operation). For a given node $j$, the aggregate is obtained by summing the features defined on the edges connected to this node (${\cal M}(j)$ in the summation subscript is the set of neighbor node indices for owner node $j$). As such, the aggregation step directly invokes the graph connectivity and provides to the message passing layer the key ability to model non-local interactions. The synchronization process in Eq.~\ref{eq:mp_sync} occurs directly on these aggregates and resembles a mean scatter operation. For a given node $j$, the set ${\cal C}(k)$ contains all node indices at the same spatial position ${\bf p}_j$ within the element graph neighborhood. As the name implies, the synchronization effectively reassigns the aggregate value to the mean of all aggregates at coincident locations in physical space. The synchronized aggregate, together with the node features from the previous layer, is then passed to the node update MLP in Eq.~\ref{eq:mp_node_update}, thereby concluding the synchronized message passing layer. 

It should be noted that after embedding the neighborhood information into the coarse-scale query graph $\widetilde{G}^q_c$ by means of the CSP, the neighborhood graph elements in ${\cal N}(G_c^q)$ are unused in successive GNN operations (all relevant neighborhood information for the super-resolution task is assumed to be contained within latent query graph at this stage). 

\subsubsection{Graph Unpooling/Interpolation Layer}
\label{sec:unpool}
The goal of the unpooling stage is to lift the embedded coarse-scale representation (the node features in the latent query graph $\widetilde{G}_c^q$) to the corresponding fine-graph representation in the same element. Such a lifting operation is evaluated within the GNN forward pass and is required to complete the element-local super-resolution procedure. {\color{black}Note that an unpooling layer is also used \textit{outside} of the core architecture to facilitate a residual-based GNN prediction (refer to Fig.~\ref{fig:gnn_arch}) -- this distinction is explained in \ref{sec:app:unpool}}

The unpooling operation can be concisely represented as $\widetilde{G}_f^q \leftarrow \text{Unpool}(\widetilde{G}_c^q)$, where $\widetilde{G}_f^q$ is the latent graph of the query element on the fine-scales (as indicated by the change in subscript). For example, in the case of P=1 to P=7 mapping, the connectivity for $\widetilde{G}_f^q$ is contained in the element-local adjacency matrix ${\bf A}_7$ (see Fig.~\ref{fig:scope_graphgen} (right) and discussion in Sec.~\ref{sec:graph_gen}). To maintain the fully graph-based nature of the architecture, the unpooling procedure is carried out using a K-nearest neighbors algorithm that recovers interpolated node features on $\widetilde{G}_f^q$ using inverse distance-weighted node features on the coarse graph $\widetilde{G}_c^q$. In a first step, for the $j$-th node position ${\bf p}_{j,f}$ on the fine graph, the set of $K$ closest nodes on the coarse graph in physical space is recovered as $\{{\bf p}_{1,c}, \ldots, {\bf p}_{K,c}\}$, where $K=8$ here. Given this set and the corresponding set of node features at the same positions, the node features at the $j$-th fine-scale graph node, denoted ${\bf y}_{j,f} \in \mathbb{R}^{N_H}$, are given by
\begin{equation}
    \label{eq:knn_interp}
    {\bf y}_{j,f} = \frac{\sum_{k=1}^K w_k {\bf y}_{k,c}}
    {\sum_{k=1}^K w_k}, \quad \text{where} \quad w_k = \frac{1}{\lVert {\bf p}_{j,f} - {\bf p}_{k,c} \rVert _2^2}.  
\end{equation}
In Eq.~\ref{eq:knn_interp}, ${\bf y}_{k,c}$ is the node feature on the coarse graph node within the nearest-neighbor set of the fine node query. This procedure is carried out for all fine graph nodes in the query element to populate their node features. Edge features are subsequently re-initialized on $\widetilde{G}_f^q$ using relative unpooled node features and positions. 

\subsubsection{Fine-Scale Processor}
\label{sec:fine_scale_processor}
The above unpooling operation allows for additional message passing layers to be executed at finer length scales in the \textit{fine-scale processor} (FSP) stage, producing a type of multiscale message passing model for sub-grid scale information recovery. More specifically, in the FSP, an additional set of $F$ message passing layers is called on the node and edge features of the unpooled graph $\widetilde{G}_f^q$. These message passing layers are equivalent to those used in the CSP described in the previous section, but without the synchronization step (this step is no longer necessary, as the neighborhood is discarded after the action of the CSP). 

Separated by the unpooling operation, the presence of coarse-scale and fine-scale message passing operations (provided by the CSP and FSP, respectively) allows one to effectively control the degree of super-resolution dependency on coarse and fine scales through prescription of the number of message passing layers, $C$ and $F$, in each stage. As demonstrated in Sec.~\ref{sec:results}, this can be leveraged to compare mesh-based super-resolution capability in the context of a purely coarse-scale model (by setting $F = 0$ and $C>0$), and a multiscale model (by setting $F>0$ and $C>0$), wherein the FSP process can be interpreted as a type of fine-scale corrector, leading to useful physical insights.  

\subsection{Scaling and Training Objective Functions}
Objective functions are evaluated on scaled element-local velocity fields. The scaling (or non-dimensionalization) is performed as a pre-processing step on the input coarse query element and neighboring element velocities before the GNN evaluation. The scaling procedure for element $i$ is given by
\begin{equation}
    \label{eq:scaling_input}
    {\bf y}_{1,i}^\text{scaled} = \frac{{\bf y}_{1,i} - \boldsymbol{\mu}_{1,i}}{\boldsymbol{\sigma}_{1,i}},
\end{equation}
where ${\bf y}_{1,i}$ is the input coarse velocity field for the query element (equivalent to the node attribute matrix for the query graph), ${\bf y}_{1,i}^\text{scaled}$ is corresponding scaled quantity, $\boldsymbol{\mu}_{1,i} \in \mathbb{R}^3$ is the mean velocity field \textit{local to the query element} $i$, and $\boldsymbol{\sigma}_{1,i} \in \mathbb{R}^3$ is the element-local standard deviation of the velocity field. Velocity fields of neighboring coarse elements (if present) are scaled using query element statistics. The target velocity fields are also scaled using the coarse query element statistics as
\begin{equation}
    \label{eq:scaling_target}
    {\bf y}_{7,i}^\text{scaled} = \frac{{\bf y}_{7,i} - \boldsymbol{\mu}_{1,i}}{\boldsymbol{\sigma}_{1,i}},
\end{equation}
where ${\bf y}_{7,i}^\text{scaled}$ denotes a P=7 velocity field local to element $i$. The local nature of the above scaling operations -- similar to the scaling procedure used in previous works \cite{karthik_vms_superres,fidkowski_sr} -- is consistent with the element-local scope of GNN operations. Although global scaling using full snapshot statistics can be utilized (removing the dependence of element index on standardization), local scaling was found to be significantly more effective for this application. In the discussion below, the distinction between scaled and unscaled velocities is omitted for brevity.

The objective function is designed to accomplish the goal of P=1 to P=7 flow-field reconstruction in a mesh-based context (i.e., a one-shot objective). In other words, upon minimizing this objective, the trained model is designed to transition an element directly from the P=1 coarse mesh to the P=7 target fine mesh in one forward pass. The objective function is given by the mean-squared error (MSE) in the velocity field as
\begin{equation}
    \label{eq:loss_oneshot}
    {\cal L} = 
    \left \langle \text{MSE}({\bf y}_{7}, \widetilde{\bf y}_{7}^1) \right \rangle, 
\end{equation}
where ${\bf y}_{7}$ is the target velocity field,  $\widetilde{\bf y}_{7}^1 = \widehat{\bf y}_{7}^1 + \widetilde{\bf r}_{7}^1$ is the GNN prediction, and  $\widehat{\bf y}_{7}^1$ is the interpolated input velocity field in the query element obtained using the same K-nearest neighbors strategy described in Sec.~\ref{sec:unpool}. In Eq.~\ref{eq:loss_oneshot}, variables correspond to three-dimensional velocity fields in an element (they are node attribute matrices and not individual node vectors). As such, the mean in the MSE evaluation is taken over all fine $N_P = (P+1)^3$ query element graph nodes and their features, which are the x, y, and z velocity components. Variable superscripts, when present, correspond to the starting element order, and subscripts to the final (or target) element order. The brackets in Eq.~\ref{eq:loss_oneshot} represent an ensemble (or batch) average over all elements in the training set. It is emphasized that, as per GNN modeling scope in Eq.~\ref{eq:srgnn}, the GNN output is the residual $\widetilde{\bf r}_{7}^1$ \textit{and not the velocity field directly}: the GNN output is added to the interpolated input via residual connection to recover the final prediction $\widetilde{{\bf y}}_7^1$ for a given element. 

{\color{black}\section{Results}}
\label{sec:results}
With the super-resolution strategy described in Sec.~\ref{sec:methodology}, the goal of this section is to demonstrate the method in various contexts. This is done in the following sections. First, in Sec.~\ref{sec:results:tgv} and Sec.~\ref{sec:results:bfs}, the GNN performance is independently analyzed on the TGV and BFS datasets, respectively, focusing on model evaluations on unseen snapshots at fixed Reynolds numbers and geometry configurations (i.e., models are trained and tested at the same Reynolds numbers in these sections to isolate capability of temporal generalization). Then, in Sec.~\ref{sec:results:reynolds_extrap}, assessment of models extrapolated to out-of-distribution Reynolds numbers is performed using the TGV configuration. Lastly, in Sec.~\ref{sec:results:cavity}, assessment of models in a geometry extrapolation context is performed using the cavity case. 

\underline{\textbf{Implementation details:}} All models are implemented using a combination of PyTorch \cite{pytorch} and PyTorch Geometric \cite{pygeom} libraries. MLPs contain two hidden layers with ELU activations \cite{elu}, layer normalization \cite{layernorm}, and residual connections. Hidden node and edge feature dimensionalities are set to 128. Each model was trained for 100 epochs, with 10\% of all samples in respective datasets set aside during training for validation purposes. The Adam optimizer \cite{adam} was used to train all models with initial learning rate set to $10^{-4}$. A plateau-based learning rate scheduler (the \verb|ReduceLROnPleateau| function) was used to dynamically adjust learning rates to mitigate effects of stagnation and increases in the validation loss. Models were trained using 8 Nvidia A100 GPUs (batch size of 4 per GPU) using two nodes of the Polaris supercomputer at the Argonne Leadership Computing Facility (ALCF). All results below are reported using predictions that have been unscaled/re-dimensionalized with respect to element-local statistics. 

\underline{\textbf{Spectral element interpolation:}} GNN predictions are juxtaposed with an interpolation operation typically used to initialize fine-scale flow-fields in multigrid cycles in the spectral element solution procedure. This is referred to as "spectral element (SE) interpolation", the methodology of which is provided in Ref.~\cite{fischer_book}. Although the SE interpolation is not explicitly designed for turbulent flow reconstruction, and is therefore not expected to compete with GNN-based reconstructions, it constitutes a useful reference point due to both its element-local nature (the SE interpolation can be interpreted as a linear prolongation operator acting on the element-local velocity fields, resembling the localized scope of the GNN operation) and the fact that it is already implemented in the \verb|NekRS| flow solver. As such, reference comparisons to SE interpolation outputs will be made throughout this section. 

Alongside assessment of the effect of coarse-element neighborhood size ${\cal N}_i$, the SRGNN architecture offers a useful capability to disentangle the effect of coarse-scale versus fine-scale message passing on the super-resolution procedure through proper specification of the number of layers used in the respective processors (see Fig.~\ref{fig:gnn_arch} and the surrounding discussion). As such, the TGV demonstrations leverage the following two model configurations:
\begin{itemize}
    \item \underline{\textbf{Model 1 -- Coarse-scale:}} In this configuration, the total number of message passing layers in the FSP is set to zero ($F=0$), and the number of layers in the CSP is set to 12 ($C=12$). The absense of fine-scale message passing therefore renders the Model 1 configuration a \textit{purely coarse-scale} model, as the only learnable non-local functions are those that model information exchange at length scales corresponding to coarse element edge lengths. 
    \item \underline{\textbf{Model 2 -- Multiscale:}} This configuration adds layers to the fine-scale processor. More specifically, the number of message passing layers is $C=6$ in the CSP is $F=6$ in the FSP. The number of layers in the CSP is dropped from 12 to 6 here to accommodate the addition of new layers in the FSP, so as to not increase the total number of parameters in the message passing process. The resulting architecture is termed \textit{multiscale} due to the presence of learnable non-local functions operating at both coarse and fine length scales. The action of the FSP in this context can be interpreted as a ``corrector" to the interpolated query element in latent space produced by purely coarse scale operations. 
\end{itemize}

{\color{black} The total number of message passing layers in both Model 1 and 2 was set to 12 based on empirical tests that demonstrated good model performance up to this value. Additionally, at this number of message passing layers, and using the message passing variant described in Sec.~\ref{sec:gnn_architecture}, over-smoothing effects are not observed \cite{meshgraphnet}. Training costs, in terms of total epoch times and model throughput extracted from TGV datasets, are shown in Table~\ref{tab:training_cost} as a reference.}

\begin{table*}[ht]
\centering
\begin{tabular}{|c|c|c|c|c|c|c|} 
\hline
\multirow{2}{*}
    {
        \begin{tabular}{c}
           GNN Configuration:\\Coarse element neighbors:
        \end{tabular}
    } & \multicolumn{3}{c|}{\textbf{Model 1} (Coarse-scale)} & \multicolumn{3}{c|}{\textbf{Model 2} (Multiscale)}\\ 
\cline{2-7} 
 & \textbf{0 Nbrs} & \textbf{6 Nbrs} & \textbf{26 Nbrs} & \textbf{0 Nbrs} & \textbf{6 Nbrs} & \textbf{26 Nbrs} \\ 
\hline
    Training time, 1 epoch [sec] & 128.2 s & 144.7 s & 151 s & 141.9 s & 150.4 s & 155.2 s \\ 
    Training throughput [elements/sec] & 982.6 & 870.6 & 834.3 & 887.8 & 837.6 & 811.7 \\
\hline
\end{tabular}
\caption{{\color{black}GNN training time and throughput (in terms of elements processed per second during training). Data collected from Re=3200 TGV training runs, using 2 Polaris nodes (8 total Nvidia A100 GPUs). Total number of training elements was 125973 (this comes from roughly 90\% of the total 139968 elements in the TGV dataset, with the remaining elements used as a validation set).}}
\label{tab:training_cost} 
\end{table*}

\subsection{Demonstration on Taylor-Green Vortex (TGV)}
\label{sec:results:tgv}

With regard to super-resolution accuracy, emphasis here is placed on comparison between the above described Model 1 and Model 2 configurations from the perspectives of instantaneous energy spectrum (global) analysis and qualitative visualizations of super-resolved flow-fields. Alongside the distinction between Model 1 and Model 2 performance, discussion is centralized around the influence of coarse element neighborhood sizes on model prediction accuracy. Some additional analysis of element-local super-resolution error is provided in \ref{sec:app:tgv_local_error}. 

\subsubsection{Global Error Analysis}
\label{sec:results:tgv:global_error}
\begin{figure}
    \centering
    \includegraphics[width=0.9\textwidth]{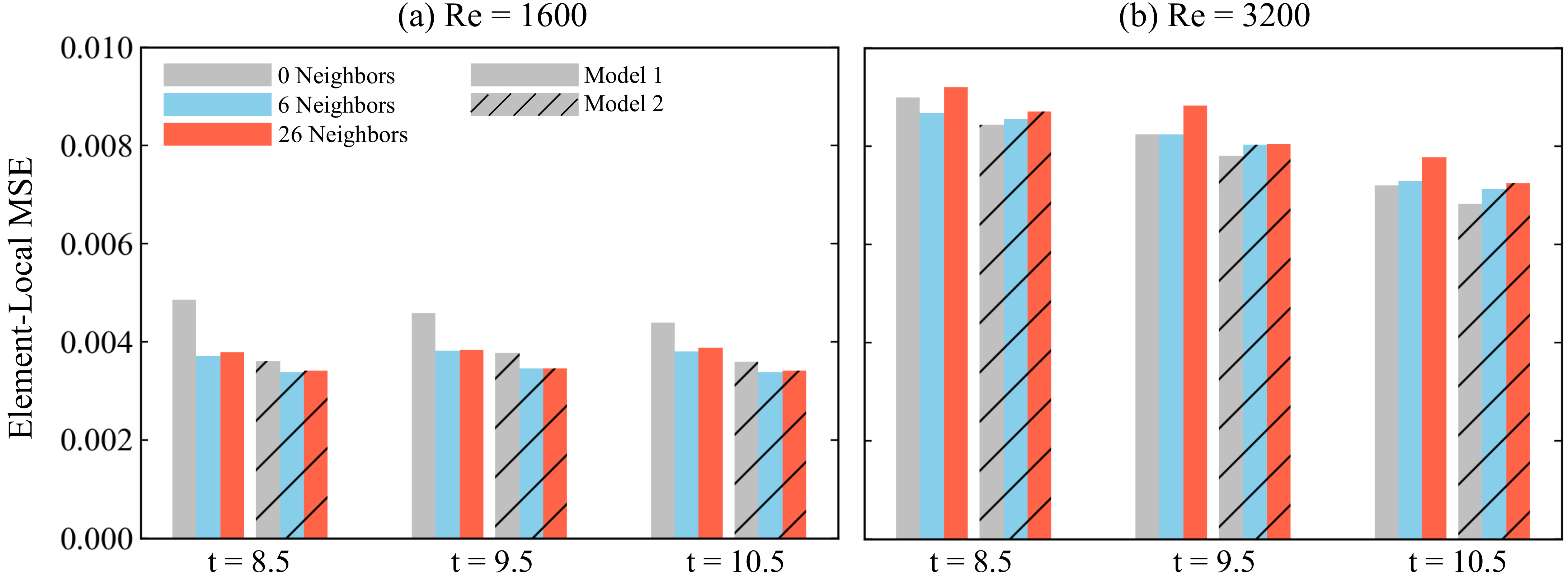}
    \caption{Query element MSE averaged over all elements in one-shot super-resolution task in test set snapshots for (a) {\color{black}TGV} models trained and evaluated using Re=1600 data and (b) {\color{black}TGV} models trained and evaluated using Re=3200 data. Each bar represents a separately trained GNN: errors shown for models using 0, 6, and 26 coarse element neighbors in both Model 1 (coarse-scale) and Model 2 (multiscale) GNN configurations.}
    \label{fig:mse_time_extrap}
\end{figure}

\underline{\textbf{Element-Local Mean-Squared Errors:}} To provide a global evaluation of the impact of model configuration, element neighborhood size, and Reynolds number on super-resolution accuracy, objective functions (in the form of average element-local mean-squared errors, as per Eq.~\ref{eq:loss_oneshot}) for the respective models evaluated on the unseen test set snapshots are shown in Fig.~\ref{fig:mse_time_extrap}(a) and (b) for models trained (and evaluated on) Re=1600 and 3200 datasets, respectively. The figure reveals an increase in test set errors from Re=1600 to 3200 by a factor of roughly 2 in all model configurations, directly proportional to the increase in Reynolds number. This increase in error can be attributed to the significant added complexity in the Re=3200 dataset in terms of fine-scale turbulence structures relative to the Re=1600 case (see \ref{sec:app:tgv}). The implication is that for a fixed model architecture (a fixed number of message passing layers), the super-resolution procedure encounters greater difficulty in direct proportion to the Reynolds number, a quality consistent with physical expectations. Additionally, at Re=1600, errors in each snapshot drop noticeably in Model 1 when moving from 0 to 6 neighbors, whereas in the Model 2 setting, less dependency is seen on the number of neighbors to the overall error.

\begin{figure}
    \centering
    \includegraphics[width=\textwidth]{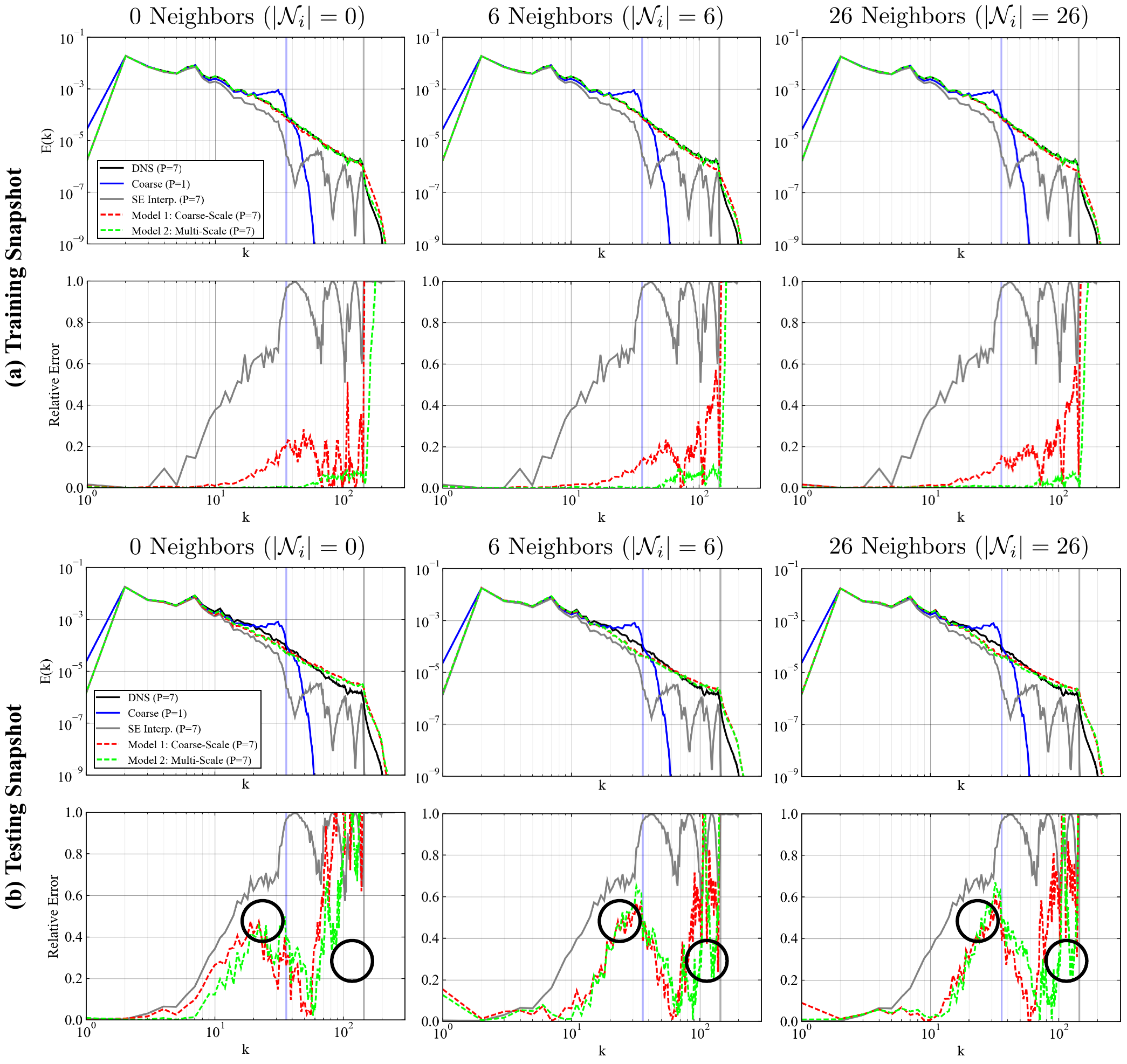}
    \caption{\textbf{(a)} \underline{Top row}: Instantaneous {\color{black}TGV} energy spectra for training set snapshot (t=10) at \textbf{Re=1600}. Curves shown for DNS target (black), coarse input (blue), spectral interpolation (gray), Model 1 (coarse-scale GNN, dashed-red), and Model 2 (multiscale GNN, dashed-green). \underline{Bottom row}: Corresponding relative errors with respect to DNS spectrum versus wavenumber. \textbf{(b)} Same as (a), but for a time-extrapolated snapshot at t=10.5 also at Re=1600. As indicated by titles, plots from left to right correspond to increasing number of coarse element neighbors used in predictions (see Fig.~\ref{fig:scope_graphgen}).}
    \label{fig:gnn_spectrum_1600}
\end{figure}

\begin{figure}
    \centering
    \includegraphics[width=\textwidth]{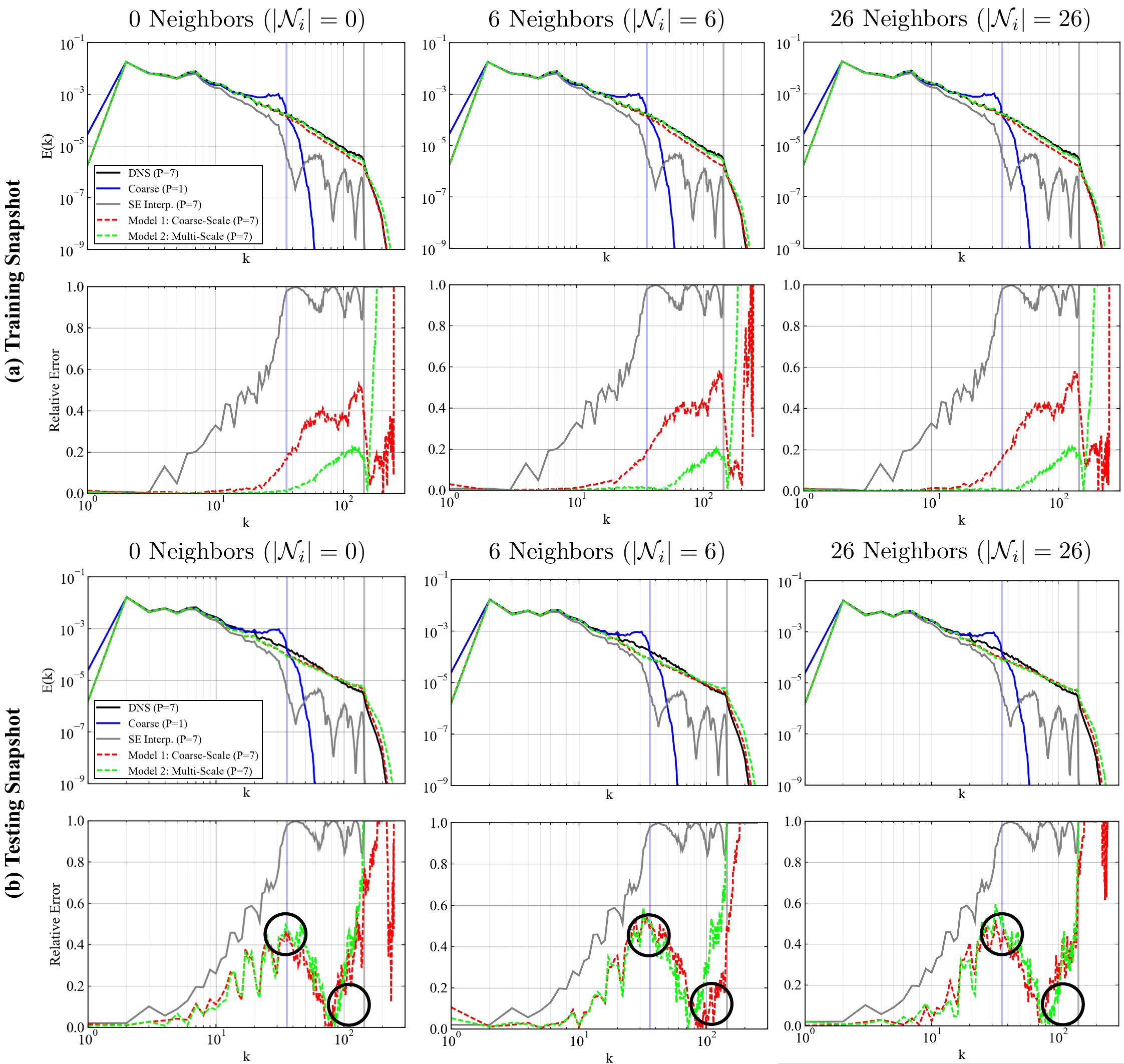}
    \caption{\textbf{(a)} \underline{Top row}: Instantaneous {\color{black}TGV} energy spectra for training set snapshot (t=10) at \textbf{Re=3200}. Curves shown for DNS target (black), coarse input (blue), spectral interpolation (gray), Model 1 (coarse-scale GNN, dashed-red), and Model 2 (multiscale GNN, dashed-green). \underline{Bottom row}: Corresponding relative errors with respect to DNS spectrum versus wavenumber. \textbf{(b)} Same as (a), but for a time-extrapolated snapshot at t=10.5 also at Re=1600. As indicated by titles, plots from left to right correspond to increasing number of coarse element neighbors used in predictions (see Fig.~\ref{fig:scope_graphgen}).}
    \label{fig:gnn_spectrum_3200}
\end{figure}

\underline{\textbf{Energy Spectra:}} To supplement the global MSE metric in Fig.~\ref{fig:mse_time_extrap}, instantaneous energy spectra and corresponding errors in spectra predictions relative to the P=7 targets are shown in Fig.~\ref{fig:gnn_spectrum_1600} and \ref{fig:gnn_spectrum_3200} for Re=1600 and 3200, respectively. Assessment of model evaluations on the training snapshots in Fig.~\ref{fig:gnn_spectrum_1600}(a) and \ref{fig:gnn_spectrum_3200}(a) serve as validation for the training procedure. Immediately apparent in both cases is that the GNN models are substantially more effective in recovering the target spectrum than the SE interpolation, which again is expected. It should be noted that the SE interpolation operator does recover some fine-scale information -- it eliminates the accumulation of energy in the coarse P=1 solution in the inertial range and transfers this energy in a dispersed fashion into the high-wavenumber regime.

Super-resolved spectra for test snapshots are shown in Fig.~\ref{fig:gnn_spectrum_1600}(b) and \ref{fig:gnn_spectrum_3200}(b) for Re=1600 and 3200, respectively; these figures illustrate the generalization capability of the models in a time-extrapolated setting. At Re=1600, inspection of the spectrum curves (top row of Fig.~\ref{fig:gnn_spectrum_1600}(b)) shows how all models retain the ability to recover energy at the high-wavenumbers, and patterns in the low-wavenumber regimes are qualitatively similar to training set results. However, in the high-wavenumber region (particularly near $k=100$), the effect of slight over-predictions of energy content relative to the target is present at all neighborhood sizes. The relative error curves (bottom row) highlight key differences between training set predictions in two facets, which are less apparent on visual inspection of spectrum curves: although errors are lower than the SE interpolation for both Model 1 and Model 2 predictions, increases to relative errors appear in low-wavenumber regimes when compared to training set counterparts. These errors are non-monotonic with respect to $k$, with characteristic spikes near both low- and high-resolution Nyquist limits ($k=36$ and $k=100$, respectively).

In the Re=3200 test set predictions (Fig.~\ref{fig:gnn_spectrum_3200}(b)), two key trends are apparent: (1) test set performance is nearly identical between Models 1 and 2, despite the observed training discrepancies shown in Fig.~\ref{fig:gnn_spectrum_3200}(a), and (2) the level of energy overshoot is \textit{smaller} in the high wavenumber regime (near k=100) than in Re=1600. This is balanced by an energy undershoot near the P=1 Nyquist limit at k=36. It is this undershoot at the relatively larger scales that leads to the overall higher global errors observed in Fig.~\ref{fig:mse_time_extrap} for Re=3200. Additionally, the relative error curves in Re=3200 are qualitatively very similar to those seen at Re=1600, which indicate the same characteristic buildup to a peak in the lower wavenumber range and a subsequent dip in the higher wavenumber range -- the difference is that these features are horizontally shifted (i.e., the peak and dip in error occur at respectively higher wavenumbers), which is an effect of evaluating the same architecture at a higher Reynolds number. Overall, at the higher Reynolds number of Re=3200, the error trends between Model 1 and 2 configurations are quite similar, leading to the useful conclusion that in some settings, a model comprising purely coarse-scale learnable operations can be used to reconstruct fine-scale quantities of interest at a level comparable with a multiscale model.

\subsubsection{Effect of Neighborhood Size}
\label{sec:results:tgv:neighborhood}
\begin{figure}
    \centering
    \includegraphics[width=\textwidth]{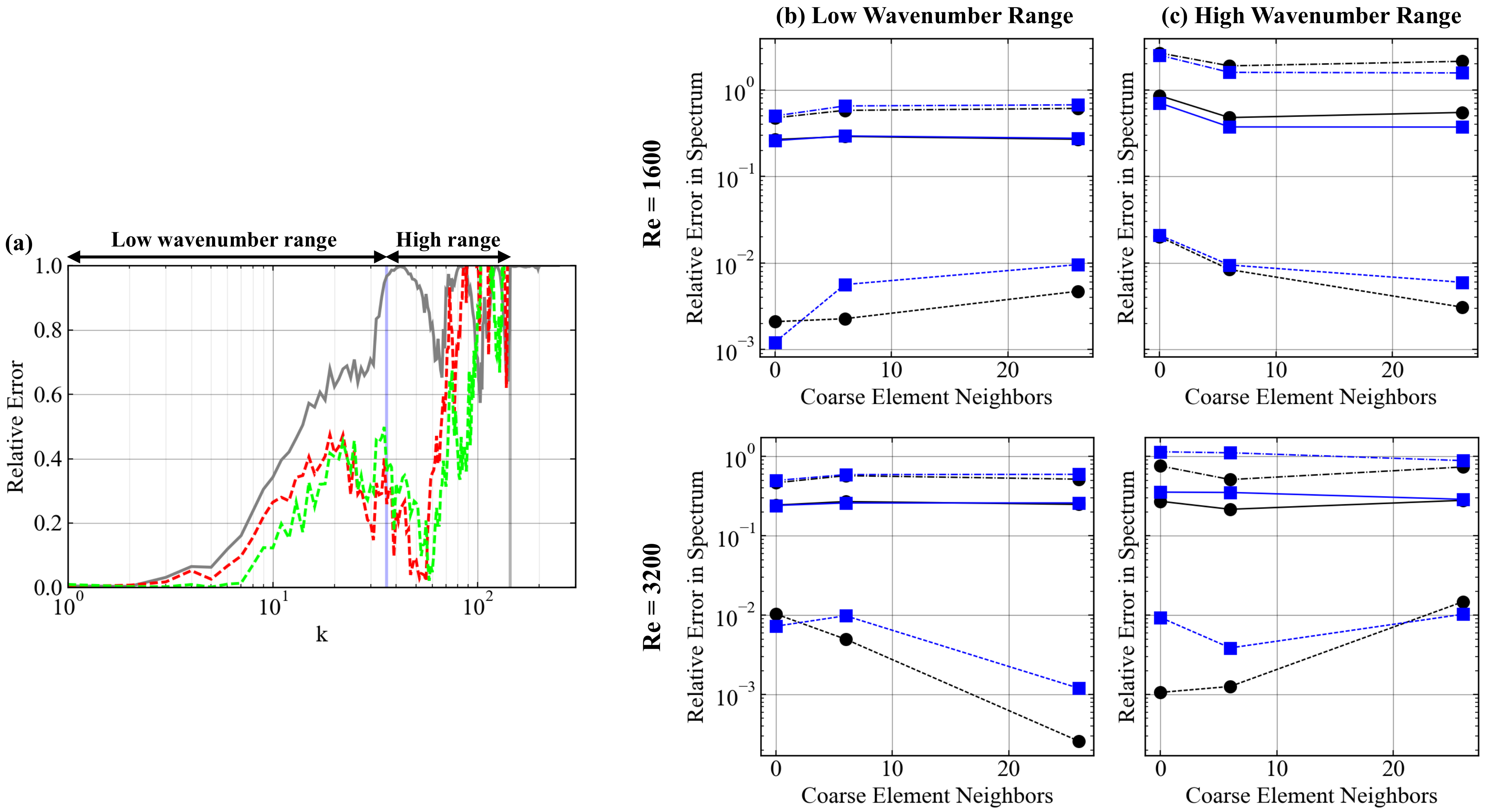}
    \caption{\textbf{(a)} Reference plot of TGV spectrum relative error versus wavenumber (reproduced from Fig.~\ref{fig:gnn_spectrum_1600}). Low and high-wavenumber ranges indicated on top of plot. \textbf{(b)} Statistics of spectrum relative errors conditioned on low wavenumber range versus number of coarse element neighbors used in GNN predictions, with Re=1600 on top and Re=3200 on bottom. \textbf{(c)} Same as (b), but conditioned on high wavenumber range. In (b) and (c), mean is solid line, maximum is dot-dashed, and minimum is dashed, with Model 1 values in black circles and Model 2 in blue squares.}
    \label{fig:gnn_error_vs_nei_1600}
\end{figure}

The effect of neighborhood size on spectrum reconstruction is provided in Fig.~\ref{fig:gnn_error_vs_nei_1600}, which plots relative error statistics versus number of coarse element neighbors in two wavenumber bins. These bins are outlined in Fig.~\ref{fig:gnn_error_vs_nei_1600}(a), and are extracted based on the P=1 and P=7 Nyquist wavenumbers shown in the vertical lines; the first bin is in the low-wavenumber range where $k=[1,36]$ (Fig.~\ref{fig:gnn_error_vs_nei_1600}(b)), and the other is in the high-wavenumber range where $k=[37,144]$ (Fig.~\ref{fig:gnn_error_vs_nei_1600}(c)). Within these wavenumber ranges, average, maximum, and minimum observed values of spectrum relative errors from Model 1 and Model 2 predictions on the test set snapshots are shown. 
When considering the low wavenumber errors in Fig.~\ref{fig:gnn_error_vs_nei_1600}(b), there is little dependency with number of coarse element neighbors on average and maximum errors at both Reynolds numbers (i.e., the neighborhood size does not play a major effect on the coarse scales). Minimum errors in the low-wavenumber regime appear to be increasing with higher neighborhood sizes in Re=1600, and decreasing in Re=3200. Although Model 1 (coarse-scale GNN) gives smaller minimum values than Model 2 in the low-wavenumber range, average and maximum value trends are nearly identical between the two. 

On the other hand, in the high-wavenumber range (Fig.~\ref{fig:gnn_error_vs_nei_1600}(c)), the effect of neighborhood size is more pronounced. Specifically, for Re=1600, increasing from 0 to 6 coarse element neighbors in both model configurations drops relative spectrum errors at high-wavenumbers across the board -- the same is not true for Re=3200, which again shows minimal sensitivity of average errors to neighborhood size. 

At Re=1600, relative errors experience no reduction in mean and maximum values beyond 6 neighbors. Unlike in the low-wavenumber range, average error curves for Model 2 -- the multiscale model -- are slightly lower than Model 1 counterparts in Fig.~\ref{fig:gnn_error_vs_nei_1600}(c). These trends ultimately reveal how, although increasing the element neighborhood sizes does not improve predictions in the low-wavenumber regime, it results in enhanced spectrum predictions at higher wavenumbers. Despite the fact that the lower average errors are observed in the higher wavenumber range for Model 2 at Re=1600, the Model 1 results are quite competitive in extrapolation settings, and even outperform Model 2 predictions in terms of spectrum error at Re=3200. Considering that Model 1 is a purely coarse-scale model, this result may seem surprising; the implication is that the message passing operations in the coarse-scale processor are able to discover non-local interactions \textit{at under-resolved length scales} in a way that allows for recovery of fine-scale information.

\subsubsection{Physical Space Visualizations}
\label{sec:results:tgv:physicalspace}
\begin{figure}
    \centering
    \includegraphics[width=0.9\textwidth]{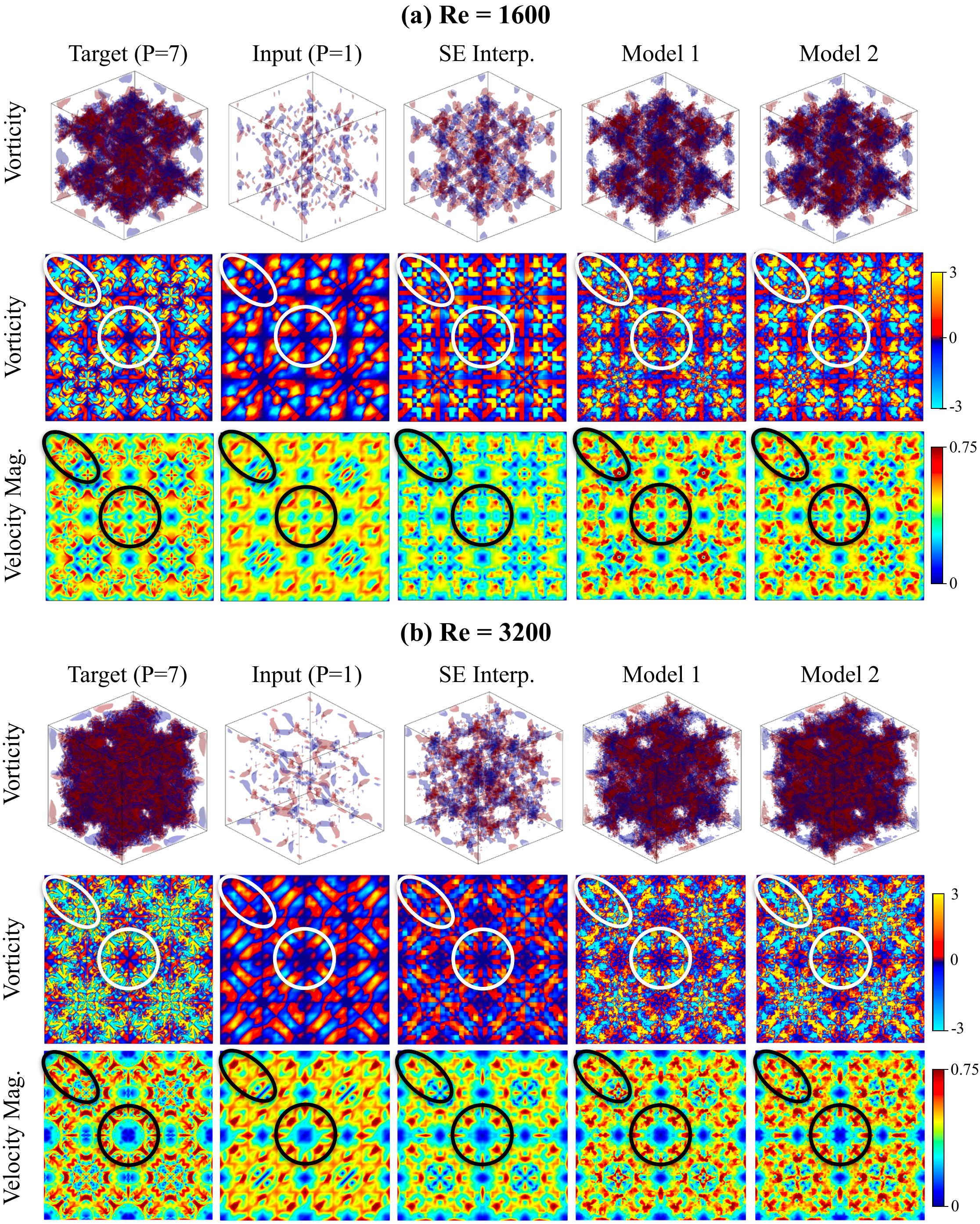}
    \caption{Instantaneous mesh-based flow-fields on time-extrapolated TGV snapshot (t=10.5s) for \textbf{(a)} Re=1600 and \textbf{(b)} Re=3200. From left-to-right: P=7 target solution (DNS), P=1 input coarse solution (projected DNS), spectral element interpolation, GNN in Model 1 configuration (coarse-scale) using 26 neighbors, GNN in Model 2 configuration (multiscale) using 26 neighbors. Top row shows z-component vorticity contours in 3D (blue and red indicate vorticity of 3 and -3 respectively), middle row shows z-component of vorticity in the X-Y plane at z=$\pi/2$, and bottom row shows velocity magnitude in the same 2D plane. Black and white circles in 2D visualizations are provided as visualization guidelines.} 
    \label{fig:physicalspace}
\end{figure}

A more qualitative assessment of the mesh-based super-resolution procedure is provided in Fig.~\ref{fig:physicalspace}(a) and (b), which shows prediction visualizations on a testing set snapshot at Re=1600 and 3200 respectively. The figure juxtaposes Model 1 and Model 2 super-resolved fields with the SE interpolation, while providing reference visualizations for target (P=7) and coarse input (P=1) fields. The vorticity contour visualizations are qualitative proxies for the turbulence intensity observed in the flow; comparison of these contours in both target P=7 and input P=1 flow-fields uncovers the inherent challenge in the super-resolution procedure, in that practically all of the turbulence is eliminated through the action of the coarse projection. Given the collection of input P=1 element graphs and their respective neighborhoods, the vorticity contour visualizations highlight the ability of GNNs to recover much of the target fine-scale structures at both tested Reynolds numbers, particularly when comparing with a conventional SE interpolation procedure. 

Although the GNNs successfully generate fine-scale flow structures, closer inspection of the super-resolved fields (best visualized in the vorticity isosurfaces near the domain boundaries in the top row plots in Fig.~\ref{fig:physicalspace}) reveals non-physical noisy artifacts in the predictions. At Re=1600, lower levels of noise are observed in the Model 2 (multiscale) flow-fields as compared to Model 1 (coarse-scale), with comparable levels of noise observed in both models at Re=3200. This mirrors the global trends in relative spectrum errors discussed in Sec.~\ref{sec:results:tgv:global_error}: the effect of these high-frequency artifacts is consistent with the presence of errors in the energy spectrum (including overshoots) observed in Figs.~\ref{fig:gnn_spectrum_1600} and \ref{fig:gnn_spectrum_3200} in the high-wavenumber regimes. 

Visualizations of vorticity and velocity magnitude fields in the X-Y planes in the middle and bottom rows of Fig.~\ref{fig:physicalspace}(a) and (b) give further demonstration of both the complexity of the turbulent flow present in the TGV configuration, as well as a more interpretable visualization of fine-scale information generation capability provided by the models. Although there are distinguishable characteristics of target flow features not present in GNN predictions, both Model 1 and Model 2 configurations generate qualitatively similar flow-fields to the target -- comparisons with both coarse-scale inputs and the SE interpolation reveal the impressive level at which fine-scale information is being injected by the GNN message passing layers. 

At Re=1600, as shown by the indicated regions in the middle and bottom rows in Fig.~\ref{fig:physicalspace}(a) (white circles), the Model 2 features appear to be more closely aligned with the corresponding target features, with further indication of elimination of many of the noisy and physically inconsistent artifacts from Model 1 predictions. It should be noted that certain multiscale features of the target flow -- such as the circular nature of the vorticity field in the middle of the X-Y plane, and its associated zone of zero vorticity flow in the direct center of the domain -- remain uncaptured in the GNN predictions. Such inconsistencies are best revealed through analysis of vorticity fields, since they depict velocity gradient differences that amplify prediction errors. At Re=3200, visualizations of these fields show significant improvement provided by the GNN models over the SE interpolation approach, albeit at a lower overall accuracy compared to Re=1600 counterparts. This aligns with the global error trends observed earlier in Fig.~\ref{fig:mse_time_extrap}. Although fine-scale features are indeed being recovered in the Re=3200 fields, the vorticity fields (and, to a lesser degree, the velocity magnitude fields) exhibit higher levels of error in the form of noisy artifacts and lower value ranges compared to the ground truth, leaving room for future improvement. 

Ultimately, the physical space visualizations in Fig.~\ref{fig:physicalspace} highlight the useful ability of the element-local GNN models to capture fine-scale information in the complex TGV configuration. Although Model 2 predictions indeed produce more accurate depictions in the case of Re=1600, the plots in Fig.~\ref{fig:physicalspace} further confirm the fact that an appreciable level of fine-scale information content can be produced by a model reliant on learning \textit{purely coarse-scale} non-local functions (Model 1), with promising reconstruction capability observed in the more challenging Re=3200 case.

{\color{black}\subsection{Demonstration on Backward-Facing Step (BFS)}}
\label{sec:results:bfs}
{\color{black}
In this section, to augment the TGV analysis, super-resolution GNNs following the \textbf{Model 2 (multiscale)} setting trained independently on the BFS data are evaluated. For model assessment, this section considers BFS-trained GNN evaluations on \textit{unseen} BFS test set snapshots at the same training Reynolds numbers (i.e., temporal extrapolation).

\subsubsection{Global Error Analysis}
Due to the non-cubic nature of the BFS domain, executing a spectrum based global error analysis in the Fourier basis is non-ideal. As a workaround, global errors are assessed using proper orthogonal decomposition (POD), which allows for the construction an orthogonal basis tailored to a specific geometry from a set of velocity field snapshots. More specifically, the POD basis vectors are extracted from target P=7 ground-truth flow-fields using \textit{the method of snapshots} \cite{sirovich1987}, which produces the basis using an eigendecomposition on the temporal correlation matrix. As this is a well-known approach, the finer details are not covered here -- the interested reader is referred to \ref{sec:app:pod} and Ref.~\cite{sirovich1987} for method details.

Instantaneous BFS POD spectra for a testing set snapshot are shown in Fig.~\ref{fig:bfs_pod_spectrum}(a) and (b) for Re=1600 and 3200, respectively. Also included in the figure is a subset of the 20 total POD modes for the Re=1600 decomposition. Apparent at both Reynolds numbers is the lack of energy content in the P=1 coarse input -- the downward shift of the P=1 energy content at all modes (blue curves in Fig.~\ref{fig:bfs_cavity_mesh}) reflects the expected elimination of all small-scale content in the flow. The relative error plots, analogous to the TGV spectrum relative errors shown earlier, highlight the GNN's capability to recover a significant amount of fine-scale information content from coarse-scale inputs. More specifically, with the exception of mode index 0 at Re=1600, GNN predictions improve on the SE interpolation baseline across the board. The benefits incurred by moving from 0 to 26 coarse element neighbors in the GNN input is especially apparent in the relative error trends. Additionally, mirroring the TGV trends, relative errors in the GNN spectra increase when moving to Re=3200, and the reconstruction advantage provided by the 26-neighbor model appears to be diminishing with increasing Reynolds number.  

Overall, the POD modes are quite difficult to interpret. However, visualization of the modes provide some avenue of interpretation for the global error reductions (or increases) provided by the GNNs at specific mode indices. Interestingly, mode index 0 is the only case at Re=1600 where the GNNs achieve higher relative error than the baseline SE interpolation. The POD mode at this index points to the activation of several \textit{large-scale} features not present in the other modes, such as the top-wall boundary layer and the shear layer emerging from the separation point. The increase in error here could be indicative of the same relative error increase at the larger length scales (smaller wavenumbers) observed in the TGV spectra (Sec.~\ref{sec:results:tgv}). Despite this, errors at all other mode indices are quite substantially reduced relative to the SE interpolation -- when observing some of the modes, it appears that these error reductions come from a more accurate reconstruction of (a) complex flow features near the reattachment point, and (b) downstream turbulence. At Re=1600, this is especially apparent for Mode 5, which shows a dramatic reduction of error when increasing the GNN coarse element neighbors from 0 to 26. This mode (alongside modes 2 and 18) contains small-scale effects near the step cavity, and highly complex small-scale features at the separation point and further downstream. 

\begin{figure}
    \centering
    \includegraphics[width=\textwidth]{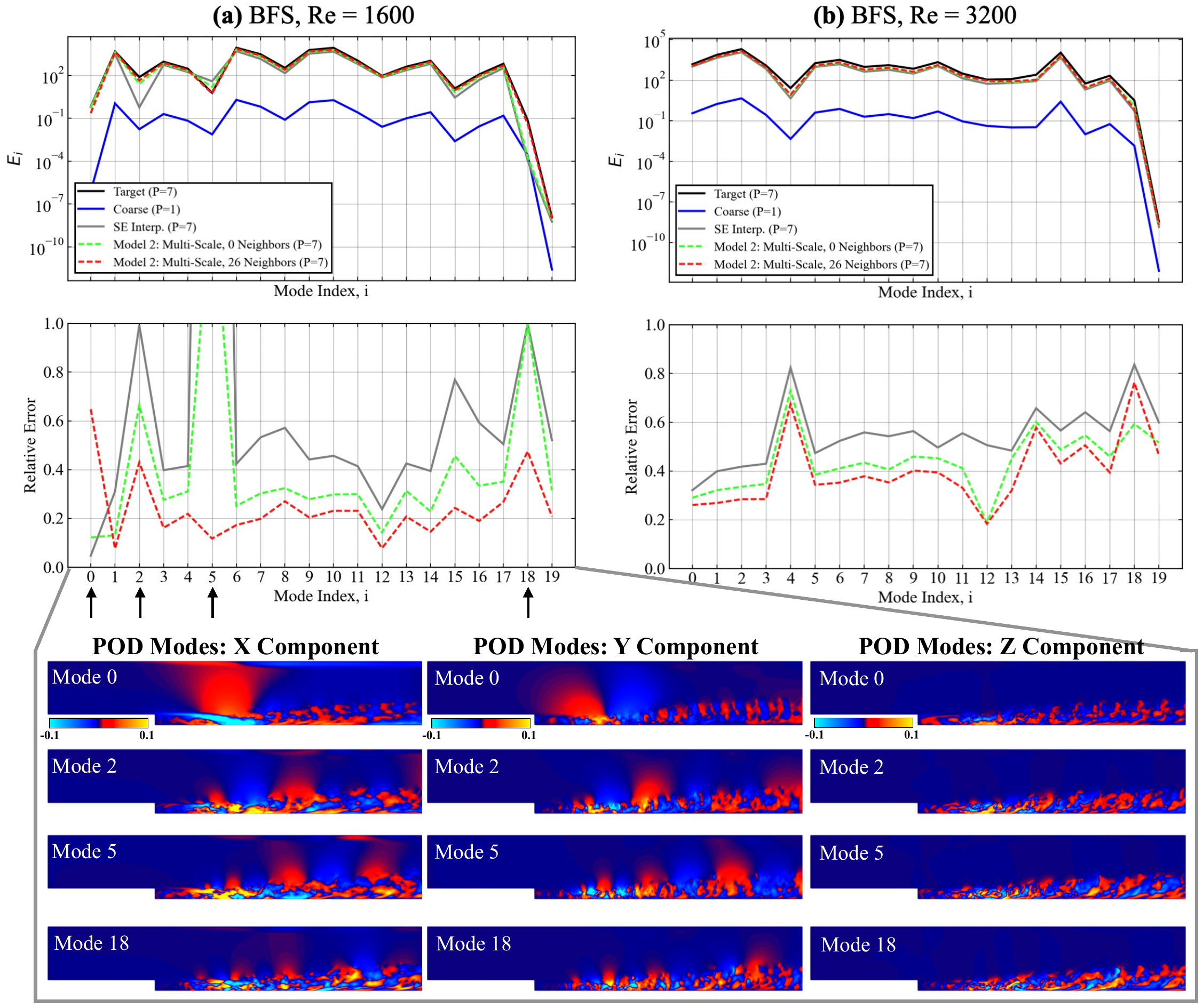}
    \caption{{\color{black}\textbf{(a)} POD energy spectrum (top plot) and relative errors in spectrum (bottom plot) evaluated at an unseen Re=1600 BFS snapshot. Curves shown for for target (black), coarse input (blue), SE interpolation (gray), GNN with 0 neighbors (dashed green), and GNN with 26 neighbors (dashed red). GNN correspond to Model 2 (multiscale) configuration. Snapshots at the bottom show visualizations of a subset of POD modes, indicated by black arrows. \textbf{(b)} Same as (a), but for Re=3200.}}
    \label{fig:bfs_pod_spectrum}
\end{figure}

\subsubsection{Physical Space Visualizations}
Physical space reconstructions for the BFS models at both Re=1600 and 3200 are provided in Figs.~\ref{fig:bfs_predictions_1} and \ref{fig:bfs_predictions_2}, which show 3d visualizations of vorticity field and additional visualizations of velocity fields on extracted 2d X-Y and Y-Z planes. 

\begin{figure}
    \centering
    \includegraphics[width=\textwidth]{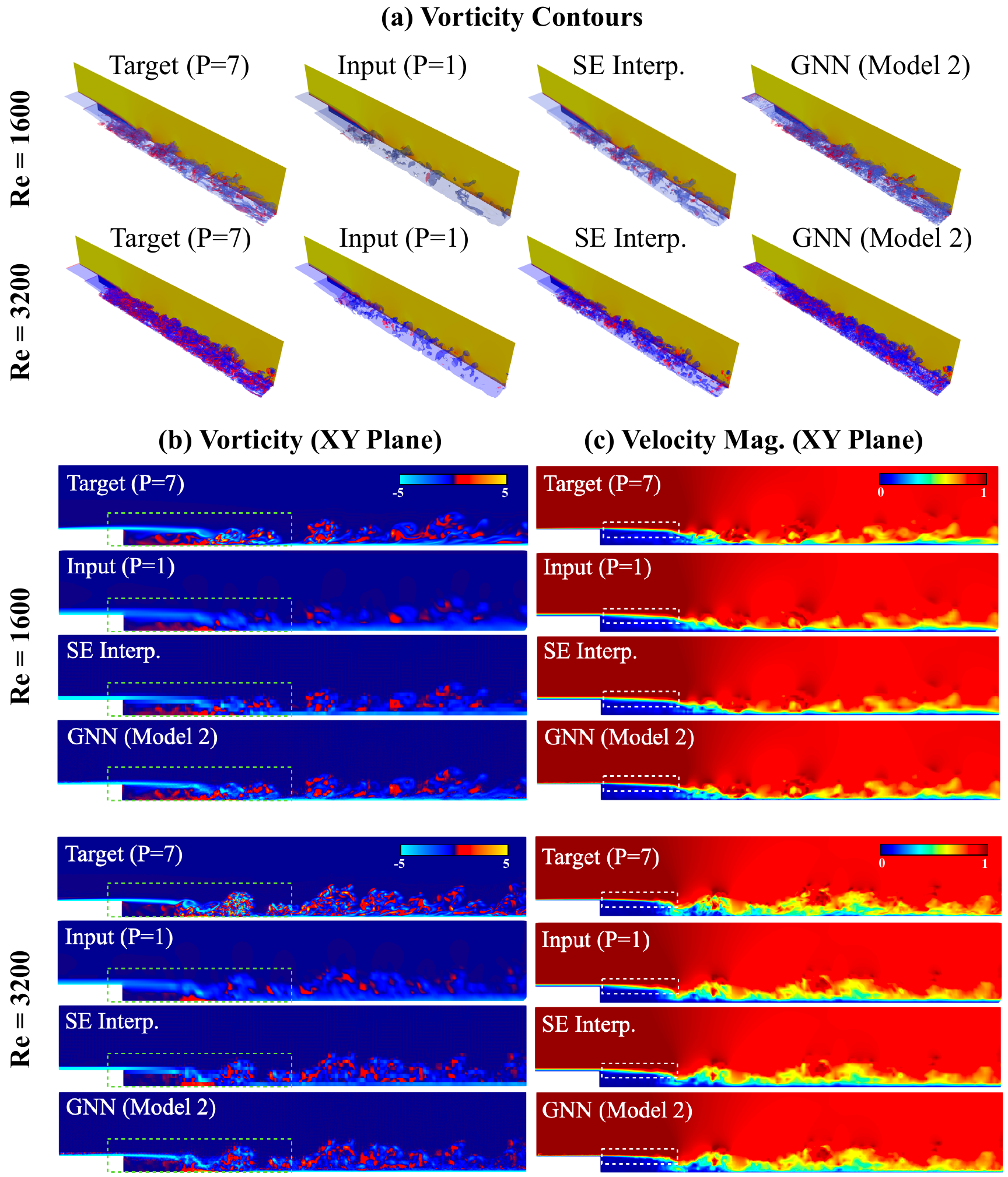}
    \caption{{\color{black}Physical space visualizations of target (P=7), input (P=1), SE interpolation, and predicted GNN (Model 2, 26 neighbors) flow-fields for the BFS at unseen Re=1600 and 3200 snapshots. \textbf{(a)} Z-component vorticity contours in 3D (blue and red indicate vorticity of 5 and -5 respectively). \textbf{(b)} Z-component vorticity fields in the XY plane (taken at z=1, the spanwise midpoint). \textbf{(c)} Same as (b), but showing velocity magnitude fields.}}
    \label{fig:bfs_predictions_1}
\end{figure}

\begin{figure}
    \centering
    \includegraphics[width=0.7\textwidth]{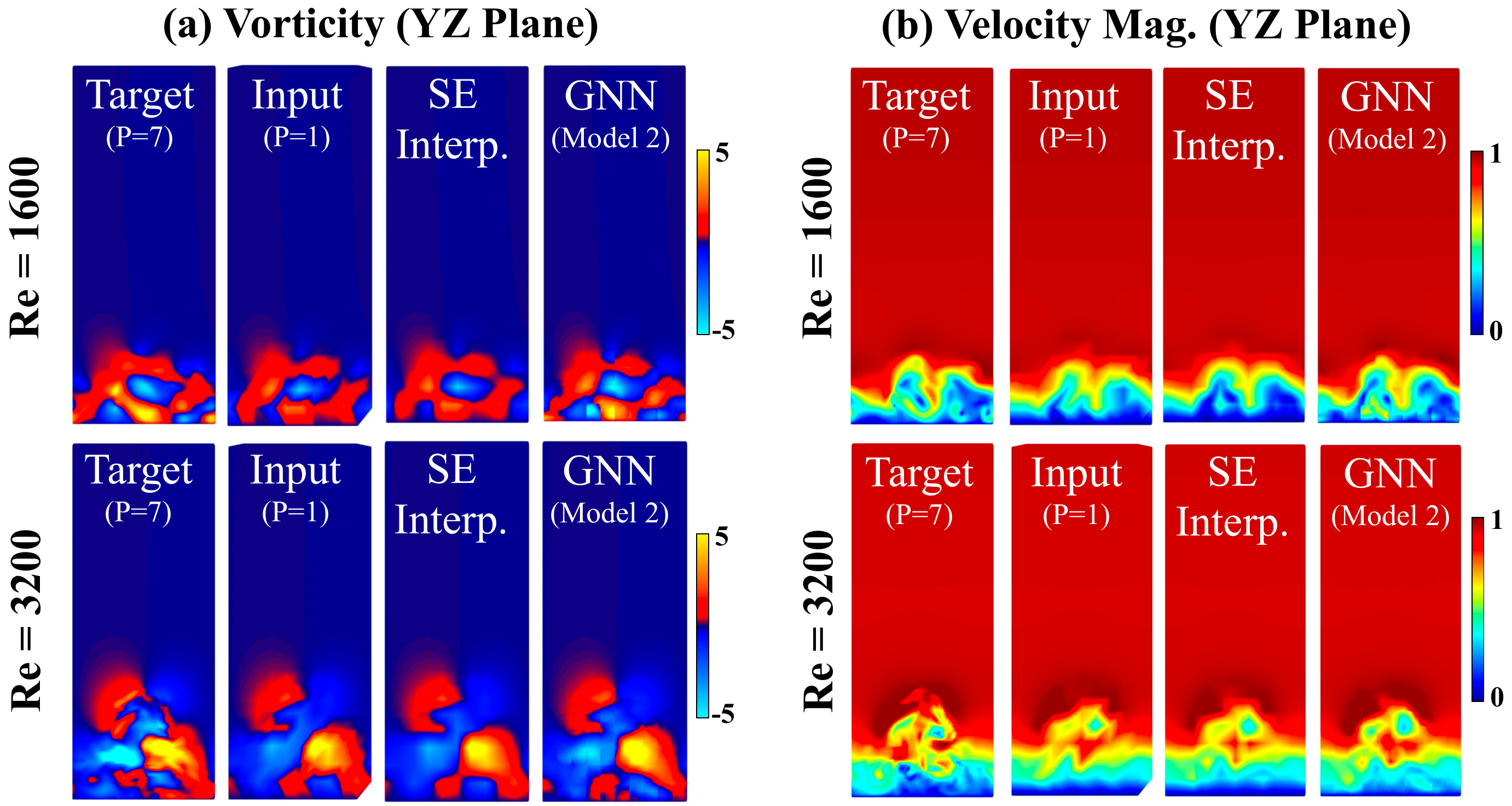}
    \caption{{\color{black}Physical space visualizations of target (P=7), input (P=1), SE interpolation, and predicted GNN (Model 2, 26 neighbors) flow-fields for the BFS at unseen Re=1600 and 3200 snapshots (same as those used in Fig.~\ref{fig:bfs_predictions_1}) \textbf{(a)} Z-component vorticity fields in the YZ plane (taken near the streamwise location of reattachment point). \textbf{(b)} Same as (a), but showing velocity magnitude fields.}}
    \label{fig:bfs_predictions_2}
\end{figure}

Inspection of the 3d vorticity contours in Fig.~\ref{fig:bfs_predictions_1}(a) shows similar qualitative behavior to the TGV trends. The GNNs (in this case, only Model 2 configurations are shown) add noticeable turbulent content to the flow from the P=1 input, resulting in expected improvement over the SE interpolation. Although the GNNs have difficulty capturing the correct vorticity contours upstream of the step anchor, the downstream turbulence enrichment is apparent, reflecting the spectrum error improvements described above. Although not shown in Fig.~\ref{fig:bfs_predictions_1}, improvement gains when jumping from 0 to 26 neighbors in the BFS GNNs come from better predictions in the upstream boundary layer, in the step cavity, and in the elimination of many nonphysical blocky artifacts in the downstream turbulence. 

A more detailed inspection of the super-resolution predictions are provided in the form of 2d planes: an X-Y plane extracted at the spanwise mid-section of the domain (shown in Fig.~\ref{fig:bfs_predictions_1}), and a Y-Z plane extracted near the streamwise location of reattachment point (shown in Fig.~\ref{fig:bfs_predictions_2}). 

Assessment of the XY planes reveal the primary streamwise-dominated flow features, such as the shear layer, separation point, and reattachment of the flow. At Re=1600, the GNN recovers a considerable amount of the vorticity content, and most crucially recovers a similar shear layer and upstream boundary layer thickness relative to the target, "unsmearing`` these quantities from the coarse P=1 input, an effect that is not captured by the SE interpolation. At Re=3200, results mirror those encountered in the TGV, suggesting a much more difficult P=1 to 7 mapping task for the GNN. Despite the fact that voriticity intensities are not captured as well as the Re=1600 case, the GNN is able to super-resolve the primary flow features fairly well in the more challenging Re=3200 scenario, eliminating many of the blocky artifacts found in the SE interpolation. A similar story is seen in the YZ planes near the reattachment point, shown in Fig.~\ref{fig:bfs_predictions_2}, which highlights recovery of some of the span-wise (3D) effects -- these planes in particular showcase the recovery of vorticity intensity and fine-scale velocity flow features in the Re=1600 case near the lower wall.
}

\subsection{Assessment of Reynolds Number Extrapolation}
\label{sec:results:reynolds_extrap}
To build on the above analysis in Sec.~\ref{sec:results:tgv} and \ref{sec:results:bfs}, which focused on purely time-extrapolation of models, the objective of this section is to investigate model performance in Reynolds extrapolated regimes. In other words, global and local error quantities are investigated with the goal of assessing how a particular GNN model trained using Re=1600 data, for example, performs when predicting super-resolved fields at Re=3200 (and vice-versa). Reynolds extrapolation is shown only for the TGV case -- general trends were found to be applicable across different configurations.

\begin{figure}
    \centering
    \includegraphics[width=\textwidth]{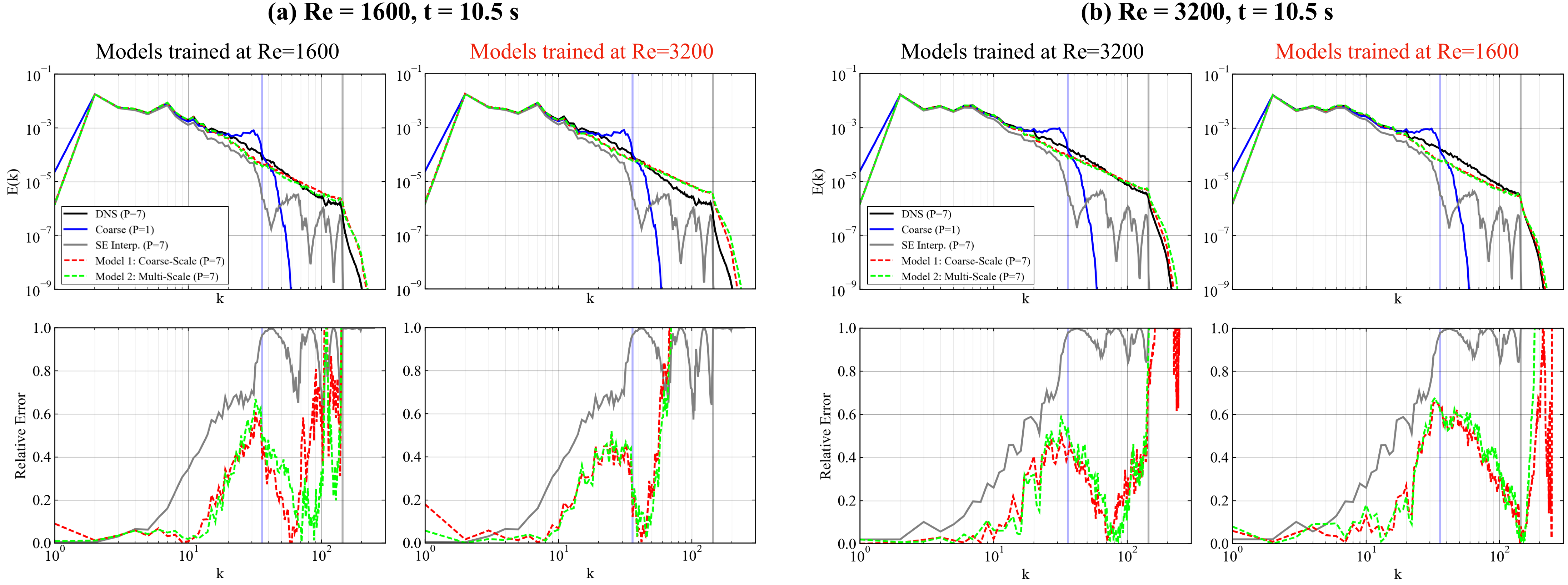}
    \caption{\textbf{(a)} Downwards-extrapolation of Reynolds number. Energy spectra for testing set snapshot (t=10.5~s) at Re=1600 (top row), alongside relative errors in spectrum (bottom row). Left column shows results using non-extrapolated model trained at the same Reynolds number (reproduced from Fig.~\ref{fig:gnn_spectrum_1600}), and right column shows results using an extrapolated model trained Re=3200. \textbf{(b)} Upwards-extrapolation of Reynolds number. Analogous to (a), but for evaluations on an Re=3200 snapshot in the upwards-extrapolation setting. In (a) and (b), GNNs correspond to the 26-neighbor setting.}
    \label{fig:spectrum_reynolds_extrap}
\end{figure}

A global perspective of Reynolds-extrapolated predictions is provided in the energy spectrum plots in Fig.~\ref{fig:spectrum_reynolds_extrap}. Figure~\ref{fig:spectrum_reynolds_extrap}(a) compares super-resolution predictions using an input P=1 field at Re=1600 for two GNNs: one trained at the same Re=1600, and the other trained using Re=3200 data. The analog is true for Fig.~\ref{fig:spectrum_reynolds_extrap}(b). As a result, Fig.~\ref{fig:spectrum_reynolds_extrap}(a) illustrates the impact of \textit{downward/reduced} Reynolds extrapolation by a factor of 2 on the predicted spectrum, and Fig.~\ref{fig:spectrum_reynolds_extrap}(b) illustrates the same, but for an \textit{upward/increased} Reynolds number extrapolation.

In the Re=1600 predictions shown in Fig.~\ref{fig:spectrum_reynolds_extrap}(a), the GNN extrapolations (corresponding to models trained at Re=3200) inject additional energy at the higher wavenumbers relative to the models trained at the same Reynolds number of 1600, resulting in higher levels of overshoot and spectrum relative errors in this regime. The notable increase in errors at the higher wavenumbers are balanced by decreases to errors in the medium-range wavenumbers. In the reverse extrapolation direction in Fig.~\ref{fig:spectrum_reynolds_extrap}(b), the upwards-extrapolated models (now corresponding to those trained at Re=1600) result in essentially the inverse trend as the extrapolations in Fig.~\ref{fig:spectrum_reynolds_extrap}(a): at Re=3200, both Models 1 and 2, relative to the non-extrapolated model, tend to produce lower energies relative to the target at larger wavenumbers. In Fig.~\ref{fig:spectrum_reynolds_extrap}(b), this results in a greater level of error at the lower P=1 Nyquist limit (k=36) than the non-extrapolated model, in direct contrast to the increase in error near the \textit{high} Nyquist limit observed in Fig.~\ref{fig:spectrum_reynolds_extrap}(a). The spectra trends in the extrapolated regimes point to a quality of the super-resolution task that is inherent to this complex one-to-many mapping problem: since models were trained only at a single Reynolds number, their outputs for a given coarse-scale input are naturally "hard-coded" to the particular Reynolds number at which they are trained on, resulting in the deviations observed in Fig.~\ref{fig:spectrum_reynolds_extrap} (i.e., it is reasonable to expect a model purely trained at Re=3200 to overshoot energy when extrapolating to Re=1600, and a model purely trained at Re=1600 is expected to undershoot energy when extrapolated to Re=3200).

In light of this implication and the deviations observed in Fig.~\ref{fig:spectrum_reynolds_extrap}, a more localized analysis of flow patterns in physical space is warranted. 

\begin{figure}
    \centering
    \includegraphics[width=0.7\textwidth]{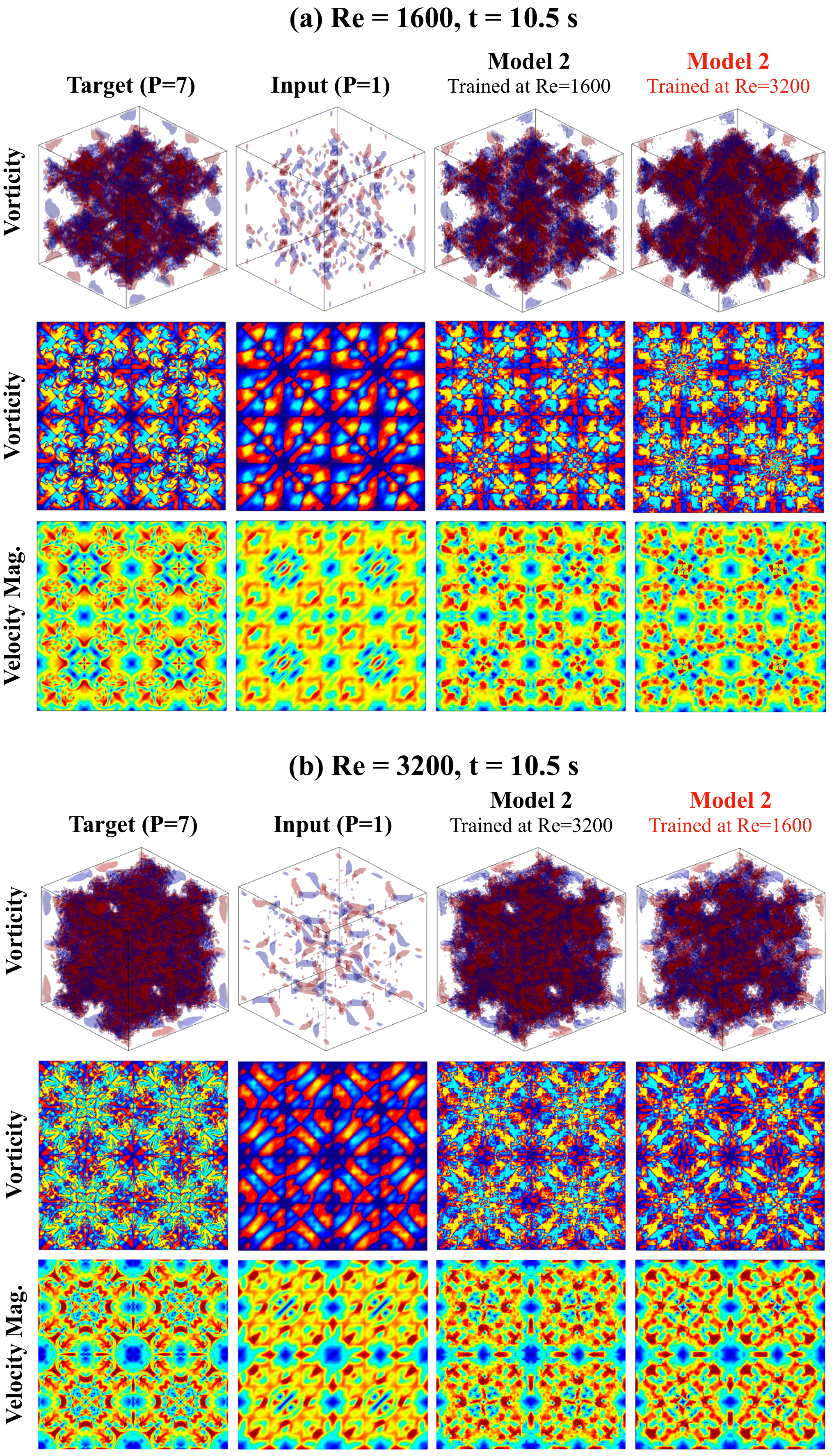}
    \caption{Physical space visualizations analogous to Fig.~\ref{fig:spectrum_reynolds_extrap}.Colorbars are consistent with those utilized in Fig.~\ref{fig:physicalspace_reynolds_extrap}. SE interpolation and Model 1 outputs are not shown for ease of visualization.}
    \label{fig:physicalspace_reynolds_extrap}
\end{figure}

To this end, Fig.~\ref{fig:physicalspace_reynolds_extrap} shows the physical space analog of Fig.~\ref{fig:spectrum_reynolds_extrap}, providing corresponding visualizations of instantaneous super-resolved flow-fields. In both Re=1600 and Re=3200 cases (Fig.~\ref{fig:physicalspace_reynolds_extrap}(a) and (b), respectively), the vorticity contour visualizations in the top rows reveal the ability of the GNN to recover much of the fine-scale turbulent structures even in extrapolated regimes. However, in the Re=1600 case (Fig.~\ref{fig:physicalspace_reynolds_extrap}(a)), since the extrapolated predictions use the model trained on Re=3200 data, vorticity contours appear more "dense" and noisy, indicative of the energy overshoot observed in Fig.~\ref{fig:spectrum_reynolds_extrap}(a) -- the opposite is true for the Re=3200 case in Fig.~\ref{fig:physicalspace_reynolds_extrap}(b), with extrapolated predictions sourced from Re=1600 models resulting in a less dense/rich (i.e., less turbulence intensity) super-resolved field relative to the target. The 2D planes of vorticity and velocity magnitude show how the downwards-extrapolation in Reynolds number (Fig.~\ref{fig:spectrum_reynolds_extrap}(a)) is in good qualitative agreement to the non-extrapolated model, and in-turn, the target. The same is not entirely true in the assessment of upwards-extrapolation in Reynolds number (Fig.~\ref{fig:spectrum_reynolds_extrap}(b)); although fine-scale structures are indeed generated by the Re=1600-trained model here, there is pronounced difficulty in producing all of the fine-scale structures as observed in the non-extrapolated model.

Ultimately, both the limitations in the upwards-extrapolation setting and the success in the downwards-extrapolation setting are reflections of the training procedure: the element-local flow-fields at Re=3200 naturally span a wider range of encountered turbulent structures relative to those observed at Re=1600, resulting in a more difficult upwards-extrapolation task in the Reynolds number. Despite this, the Reynolds extrapolations in both settings are quite promising, and augmentations to the training procedure (i.e., including additional mesh-based flow-fields sourced from more Reynolds numbers) is a promising pathway by which upwards-extrapolation results can be improved. 

\begin{figure}
    \centering
    \includegraphics[width=\textwidth]{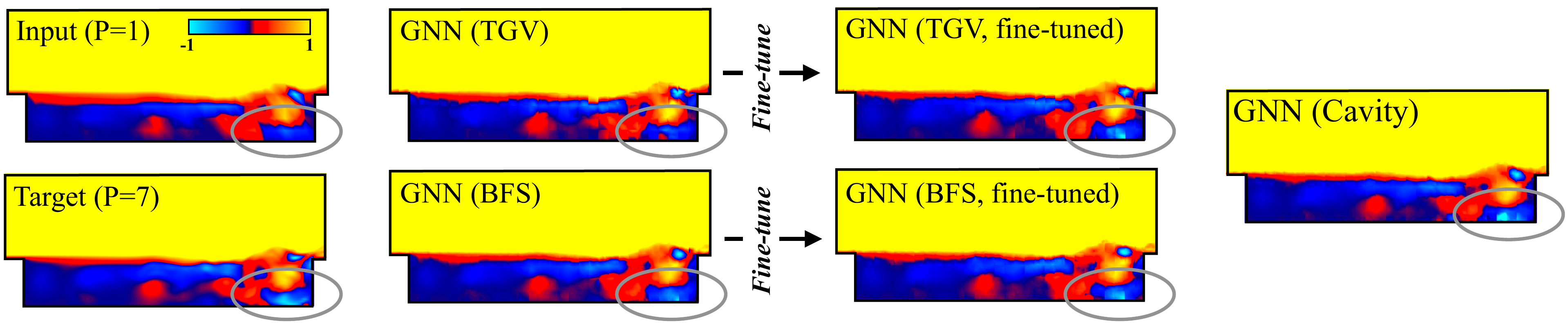}
    \caption{Visualizations of streamwise velocity component in the cavity configuration (XY mid-planes cropped near the cavity are shown) for an unseen snapshot at Re=1600, showing the effect of fine-tuning the BFS- and TGV-trained models on the cavity. All GNN models correspond to Model 2, 26-neighbor configurations. Gray circles highlight opposing cavity region, where fine-tuning improvement is observed.}
    \label{fig:cavity_finetune}
\end{figure}

{\color{black}\subsection{Assessment of Geometry Extrapolation on the Cavity}}
\label{sec:results:cavity}
{\color{black}
In this section, the cavity configuration is used to assess the zero-shot \textit{geometry} extrapolation capability of models. More specifically, in the text below, GNNs trained on both TGV and BFS configurations are used to super-resolve flow snapshots in an entirely different geometric configuration, namely the cavity flow. These evaluations are then compared to a model trained separately on cavity flow snapshots, which provides a performance upper bound on the TGV- and BFS-trained GNNs. To isolate purely geometry extrapolation behavior and errors, geometry extrapolation analysis is restricted to Re=1600.

Instantaneous cavity fields at Re=1600 are shown in Fig.~\ref{fig:cavity_vort}, which provides XY planes overlaid with 3D vorticity contours (a view into the plane) of target, input, SE interpolation, and GNN-predicted flow-fields. The GNN predictions shown in Fig.~\ref{fig:cavity_vort} correspond to the Model 2 configuration. Similar to the BFS vorticity contour plots, the visualizations showcase the ability of the GNNs to generate much of the turbulence structure in the flow. This is apparent when comparing to the SE interp outputs, which showcase highly blocky and coarse artifacts, upon which the GNNs improve. 

\begin{figure}
    \centering
    \includegraphics[width=0.7\textwidth]{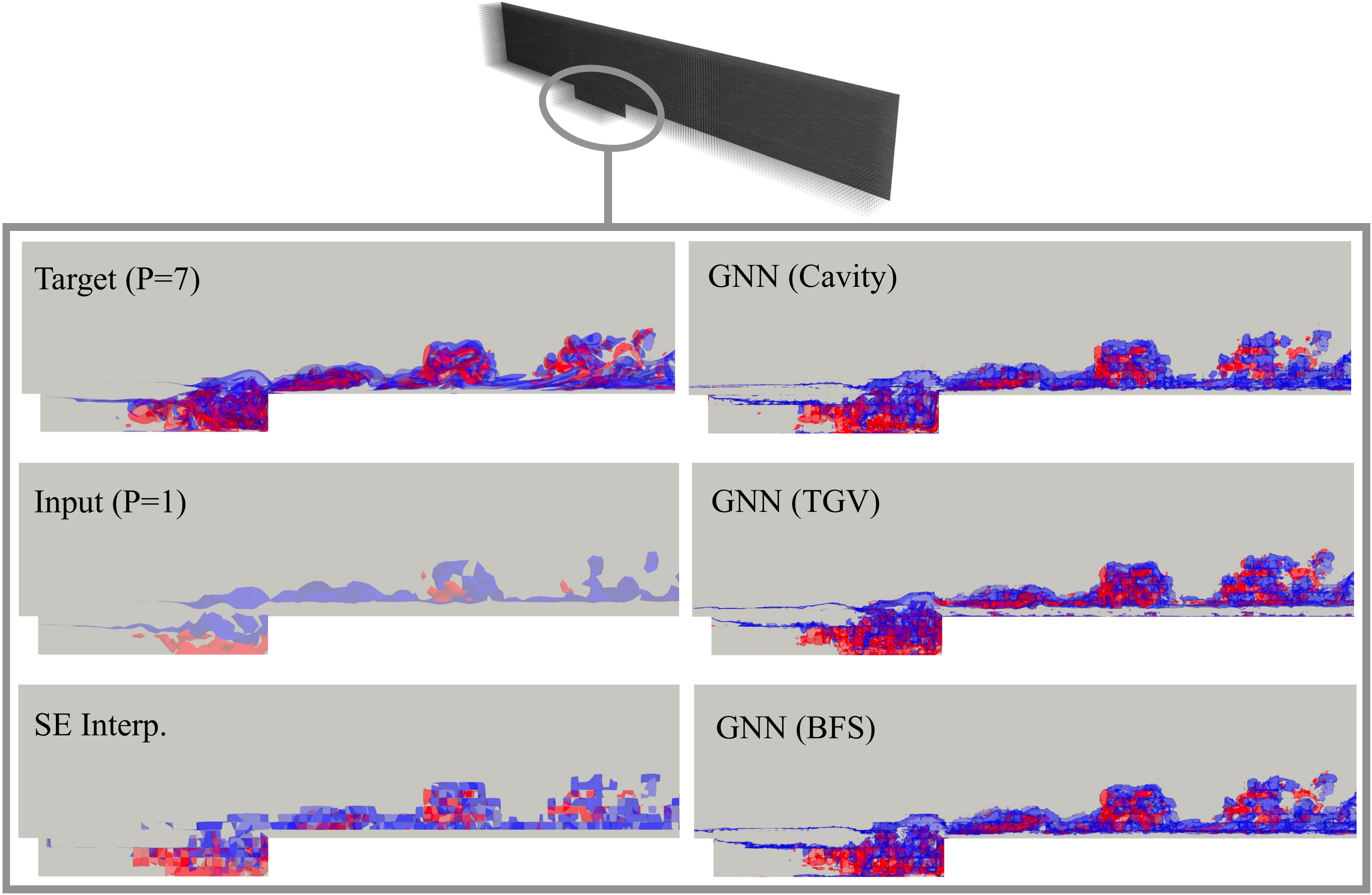}
    \caption{Physical space visualizations of vorticity contours (z-component) on an unseen cavity snapshot at Re=1600 (with zoom-in near the cavity region). Shown is target (P=7, top left), coarse input (P=1, middle left), SE interpolation (bottom left), a cavity-trained GNN prediction (top right), a TGV-trained GNN prediction (middle right), and a BFS-trained GNN prediction (bottom right). Blue and red indicate vorticity of 5 and -5, respectively. GNNs correspond to Model 2, 26 neighbor configurations.}
    \label{fig:cavity_vort}
\end{figure}

Both TGV and BFS-trained models achieve impressive reconstruction capability on the cavity configuration, although there are some deviations (i.e., the contour "streak`` just above the cavity region is not present in the target). More specifically, when looking at the vorticity contours alone, there is little qualitative difference between the cavity-trained model (the GNN baseline here) and BFS-trained model -- although the BFS and cavity configurations are indeed similar, and the BFS-trained GNN has knowledge of separation dynamics and boundary layer effects, the BFS-trained model is able to recover fine-scales near the opposing cavity wall, an aspect of the flow that was unseen during training. Perhaps most interestingly, the TGV reconstructions appear to follow the target (and cavity-trained model) quite well. This is a much more challenging extrapolation task, since the TGV-trained model has effectively no knowledge of wall and separation effects. This notion is reaffirmed when observing the presence of vorticity contours at the lower wall downstream of the cavity in the TGV-trained model, a feature that is not present in either of the BFS or cavity-trained models (as well as the target).

{\color{black}\subsubsection{Fine-Tuning on the Cavity}}
A natural extension to the above geometry extrapolation analysis is to investigate the degree to which fine-tuning can improve the prediction errors of BFS- and TGV-trained models on the cavity configuration. 

This strategy is briefly explored here in the context of the multiscale GNN (Model 2) configuration. To investigate fine-tuning, a subset of the total parameters in either the BFS or TGV models GNN are further trained (i.e., fine-tuned) on smaller cavity dataset, while all other parameters are kept frozen. Here, the parameter subset that is fine-tuned comes from the coarse-scale processor (CSP), which contains all of the coarse-scale message passing layers, while keeping the parameters in the fine-scale processor, encoder, and decoder layers frozen. This specific fine-tuning strategy is motivated by the universality of turbulent structures: the hypothesis is similar to conventional physics-based turbulence modeling intuition, in that the large-scale changes to the flow due to a configuration change require modification of the coarse-scale message passing interactions alone, and not those modeled by message passing at the finer P=7 edge lengthscales.

The effect of fine-tuning is highlighted in Fig.~\ref{fig:cavity_finetune}, which showcases streamwise velocity fields in the XY plane (zoomed into the cavity). The velocity fields also reveal additional detail into the zero-shot (not fine-tuned) model performance near the cavity -- it is evident that the TGV model observes increased non-physical discontinuities near the separation point and shear layer. Interestingly, fine-tuning, which is carried by training each model for an additional 50 epochs on a smaller dataset of 11480 cavity flow-field elements (roughly 11\% of the dataset size used to train the reference cavity GNN model), eliminates some of these non-physical artifacts in the TGV-based model. For both TGV and BFS models, fine-tuning serves to improve predictions near the rear cavity wall (gray circles in Fig.~\ref{fig:cavity_finetune}, which intuitively reflects the fact that the fine-tuning improvements are primarily felt in physical regimes that are not captured in the originating configurations (TGV and BFS). 
}

\section{Conclusion}
\label{sec:conclusion}

The goal of this work is to enable mesh-based super-resolution of fluid flows in 3D spatial configurations. A new graph neural network architecture is introduced to this end, which is designed to take as input a graph representation of a highly coarsened unresolved flow-field, and produce as output the corresponding super-resolved flow-field on a high-resolution graph in which fine-scale features exist. The graph generation procedure coincides with an element-local interpretation of flow-fields; it is consistent with strategies used in spectral element discretizations, and shares similarities with those used in finite element and finite volume discretizations. In this framework, the GNN is designed to operate not on the full mesh-based field at once, but on localized meshes of elements (or cells) directly. To facilitate super-resolution over element-local graphs, the architecture is composed of two key stages. In the first stage, a query element alongside a set number of neighboring coarse elements are encoded into a single latent graph using coarse-scale synchronized message passing over the element neighborhood (termed the coarse-scale processor). {\color{black} In the second stage, fine-scale message passing is carried out on the unpooled latent graph (termed the fine-scale processor).}  

{\color{black} It is emphasized that this GNN framework embeds certain inductive biases into the model. These include the use of GNN message passing, which inherently enforces the principle that nodes in close physical proximity exert a greater influence than those farther apart (i.e., leveraging strong spatial correlations), as well as the localization of message passing to element-local regions, introducing an additional level of locality in the model structure.}

{\color{black} Demonstration studies are performed wherein the GNN architecture is applied to achieve super-resolution in the Taylor-Green Vortex (TGV) and backward-facing step (BFS) configurations at Re=1600 and Re=3200, with hexahedral mesh-based data generated using the NekRS flow solver. Within this application, the goal is to move from a P=1 element discretization (highly coarsened projected DNS) to a P=7 (DNS-level) discretization. Additional geometry extrapolation and fine-tuning tests are conducted using a separate cavity configuration, to better assess the potential of TGV- and BFS-trained model applicability across different geometries.} 

Demonstrations of TGV super-resolution focused on performance assessments from two angles: (1) comparison of a purely coarse-scale model configuration (Model 1, with 0 message passing layers in the FSP) with a multiscale configuration (Model 2, with layers added to the FSP), and (2) the effect of the input coarse element neighborhood size on the super-resolution accuracy. Although all configurations were found to competitively achieve super-resolution in time-extrapolated scenarios (with significant improvements obtained over spectral element interpolations), model errors were found to scale roughly in proportion to increase in Reynolds number. For the TGV, at Re=1600, jumping from 0 to 6 (and 26) neighbors improved results, with Model 2 achieving generally superior reconstruction quality over Model 1 counterparts. At Re=3200, performance between Models 1 and 2 were comparable, with minimal sensitivity to the number of neighbors (i.e., at Re=3200, information contained in the surrounding element neighborhood did not play much of a role in the super-resolution process). {\color{black}In contrast, the BFS showed noticeable improvement due to an increaase in coarse element neighborhood size at both Reynolds numbers of 1600 and 3200.} 

Two Reynolds-extrapolation scenarios were tested: downwards-extrapolation (application of a model trained at Re=3200 on Re=1600 data) and upwards-extrapolation (application of a model trained at Re=1600 on Re=3200 data). Although models retained their ability to generate fine-scale information at an appreciable level, upwards-extrapolation presented a greater challenge for the GNN models, with models performing better in downwards-extrapolation settings (although downwards Reynolds extrapolation resulted in overprediction of energy at the fine scales). This is consistent with physical expectations, and was deemed to be a consequence of the training procedure: the element-local flow-fields at Re=3200 naturally span a wider range of encountered turbulent structures relative to those observed at Re=1600, resulting in a highly challenging upwards-extrapolation task in the Reynolds number.

{\color{black}Regarding geometry extrapolations, TGV- and BFS-trained models were evaluated on unseen snapshots sourced from flow over a cavity configuration. Both models were shown to achieve good reconstruction ability on the cavity, with (as physically expected) the BFS-trained models performing better than the TGV-trained ones. Fine-tuning the coarse-scale mesage passing layers on a smaller cavity dataset showed the potential for error reduction through transfer learning strategies. Although further investigation is ultimately needed to assess full extrapolation capability and fine-tuning advantages, these demonstrations highlight (a) the applicability of the proposed GNN modeling framework in handling entirely unseen mesh-based configurations, and (b) the feasibility of multi-configuration utilization of trained models. It is emphasized that, although not explored in this work, a super-resolution GNN model can be trained on element neighborhoods sourced from \textit{many} different configurations on different meshes. Such multi-configuration training is an exciting direction for developing even more robust models.}

This study presents many more avenues for future work that can take direct advantage of the GNN architecture introduced here. As mentioned in Sec.~\ref{sec:results:reynolds_extrap}, a promising extension to improve Reynolds number extrapolations is to source the training data from many different Reynolds numbers. {\color{black}Another pathway is to extend an incremental training procedure (explored to good effect in Ref.~\cite{fidkowski_sr}) to the GNN setting, which decomposes the one-shot approach into a sequence of smaller, but easier, super-resolution increments. The advantage of such an approach is that the \textit{same} GNN model can be applied at many different resolution increments, potentially providing a more robust model. This is related to the general study of mesh-agnostic behavior in GNN architectures -- on top of extending the same model to different GNN configurations, enhancing training datasets to include multiple mesh resolutions (such as in the incremental approach) is a promising route.} Additionally, within the one-shot context used here, investigation of the effect of model hyperparameters on generalization capability (such as the hidden channel dimensionality and number of message passing layers), {\color{black} as well as the role of the unpooling layer}, can be explored in greater detail. {\color{black}In particular, the role of the graph unpooling operation leveraged here, which effectively translates coarse element neighborhood latent representations to fine-scale structures, can be further investigated, along with alternate mesh-aware finite element graph lifting strategies \cite{jaiman_fem_gnn} that may improve spectral recovery.} {\color{black}Lastly, avenues for establishing enhanced theoretical understanding of the connection between GNN-based message passing and the super-resolution goal should be explored.}

\section{Acknowledgements}
The manuscript has been created by UChicago Argonne, LLC, Operator of Argonne National Laboratory (Argonne). The U.S. Government retains for itself, and others acting on its behalf, a paid-up nonexclusive, irrevocable world-wide license in said article to reproduce, prepare derivative works, distribute copies to the public, and perform publicly and display publicly, by or on behalf of the Government. This work was supported by the U.S. Department of Energy (DOE), Office of Science under contract DE-AC02-06CH11357. SB and PP acknowledge laboratory-directed research and development (LDRD) funding support from Argonne's Advanced Energy Technologies (AET) directorate through the Advanced Energy Technology and Security (AETS) Fellowship. RM acknowledges funding support from DOE Advanced Scientific Computing Research (ASCR) program through DOE-FOA-2493 project titled “Data-intensive scientific machine learning”. RBK acknowledges support by the Office of Science, U.S. Department of Energy, under contract DE-AC02-06CH11357. This research used resources of the Argonne Leadership Computing Facility (ALCF), which is a U.S. Department of Energy Office of Science User Facility operated under contract DE-AC02-06CH11357.

\bibliography{refs}

\appendix 

\section{Description of the TGV Flow}
\label{sec:app:tgv}
{\color{black}
Simulation of the TGV proceeds by solving the Navier-Stokes equations using the following initial conditions on a periodic cubic domain of length $2\pi$. Given ${\bf u}({\bf x}, t) = [u_x({\bf x},t), u_y({\bf x},t), u_z({\bf x},t)]$ and ${\bf x} = [x,y,z]$, these initial conditions are
\begin{align}
    u_x ({\bf x},0) &= \text{sin}(x)\text{cos}(y)\text{cos}(z), \\
    u_y ({\bf x},0) &= -\text{cos}(x) \text{sin}(y) \text{cos}(z), \\
    u_z ({\bf x},0) &= 0,
\end{align}
with $x,y,z \in [-\pi, \pi]$, satisfying periodic boundary conditions. The initial condition (shown in Fig.~\ref{fig:ic_mesh_element}(a)) is effectively two-dimensional -- it is characterized completely by large length scales through a single low-frequency Fourier mode, as visualized by the vorticity contours.

Despite the simplicity of the initial condition, the evolution of the TGV flow -- represented by the time evolution of the fine-scale DNS (target) flow-field ${\bf Y}_7(t)$ -- is characterized by high levels of vortex stretching and continual emergence of small-scale flow features. In contrast to forced homogeneous isotropic turbulence, velocity fluctuation statistics in high-Re TGV simulations are known to be non-stationary and transient, with velocity fields at early times exhibiting highly anisotropic behavior. This characteristic transience is evident in Fig.~\ref{fig:dissipation} (left), which plots the turbulent kinetic energy dissipation rate ($\epsilon$) versus time for $\text{Re}=1600$ and $3200$. The dissipation rate -- proportional to the spatially-integrated enstrophy and equivalent to the rate-of-change of turbulent kinetic energy \cite{diosady2015case} -- is a proxy for the strength of nonlinear vortex stretching observed in the flow. The dissipation rate for both Reynolds numbers peaks close to $t=9$ after a slow buildup, with subsequent decay at $t>9$. Consistent with results reported in Ref.~\cite{brachet1983small}, the increase in Reynolds number results in a proportional increase in the peak dissipation rate. This increase is reflected in the flow visualizations in Fig.~\ref{fig:dissipation} (right), which shows how (a) flow characteristics are both large-scale and insensitive to Reynolds number at early times, and (b) increasing Re from $1600$ to $3200$ results in enrichment of fine-scale features at times near the peak dissipation rate and beyond. 

\begin{figure}
    \centering
    \includegraphics[width=0.7\columnwidth]{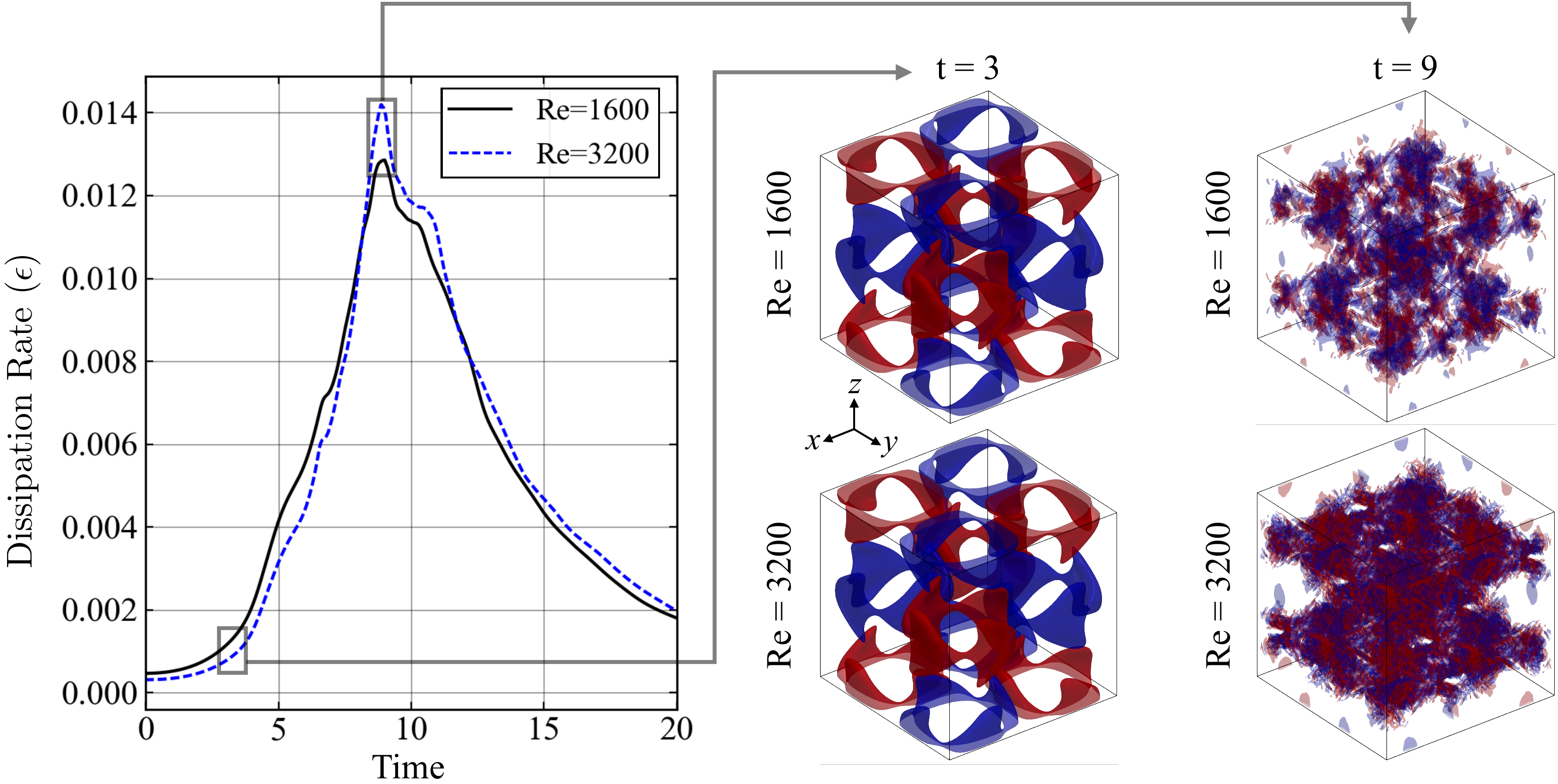}
    \caption{{\color{black}\textbf{(Left)} Temporal evolution of turbulent kinetic energy dissipation rate versus time in the TGV for Re=1600 (solid black) and Re=3200 (dashed blue). \textbf{(Right)} Visualizations of flow-fields through contours of vorticity $w_z$ at $t=3$ and $t=9$. The contours at $t=3$ reflect $w_z = 1$ and $-1$, whereas the contours at $t=9$ reflect $w_z=5$ and $-5$.}}
    \label{fig:dissipation}
\end{figure}
}

\section{Local Error Analysis of TGV Models}
\label{sec:app:tgv_local_error}
{\color{black}
Additional physical insights related to model performance can be gained through error analysis at a more local level. In this vein, Fig.~\ref{fig:std_vs_error} provides a comparison of Model 1 and 2 element-local super-resolution error for the TGV configuration (SE interpolation errors are overall higher than the GNNs, and are therefore not shown in Fig.~\ref{fig:std_vs_error} for clarity). To expose underlying relationships between coarse-scale velocity field statistics and GNN prediction errors, the figure provides velocity standard deviations in the input query element (y-axis, a coarse scale statistic) versus super-resolution prediction errors (x-axis, a fine scale statistic) at the element-local level, for both Re=1600 and Re=3200 test data. 

Apparent in Fig.~\ref{fig:std_vs_error} at both Reynolds numbers is the correlation between velocity standard deviations in the coarse input and the super-resolution error. This trend is similar for both Model 1 and Model 2 configurations at all tested neighborhood sizes. Within this trend, the shared characteristics of all models include (a) the tendency for high super-resolution errors to be correlated with high variation in coarse-scale velocities in the input element graph, and (b) a convergence in the conditional distribution of element-local MSE at higher standard deviations (i.e., the average MSEs appear to stagnate beyond standard deviation values of 0.25). Although further investigation is required for separate flow configurations, such relationships between super-resolution error and coarse-scale turbulent velocity field statistics point to promising avenues for a-posteriori error estimation at the element-local level.

At Re=1600, although the qualitative trends in Fig.~\ref{fig:std_vs_error} (top row) are similar between coarse and multiscale GNNs, the multiscale Model 2 configuration achieves consistently lower element-local errors at essentially all ranges of input velocity field standard deviations. Additionally, the plots reveal that the reduction of error when moving from neighborhood sizes above 0 is present in elements containing high levels of coarse-scale velocity variation (input standard deviations greater than 0.25), alluding to a physical connection between coarse-scale flow statistics and the contribution of coarse element neighbors to the super-resolution procedure. Such reductions in error with neighborhood size are not apparent in the Re=3200 case. However, for both Reynolds numbers, the largest discrepancies between Model 1 and Model 2 errors (up to an order of magnitude) at all neighborhood sizes are observed at the lower ranges of input element standard deviations, as shown in the inset in the rightmost plots in Fig.~\ref{fig:std_vs_error}. This indicates that the addition of fine-scale message passing layers provided by Model 2 is advantageous when a small amount of variation in input element velocity field is available at the coarse scales.

\begin{figure}
    \centering
    \includegraphics[width=\textwidth]{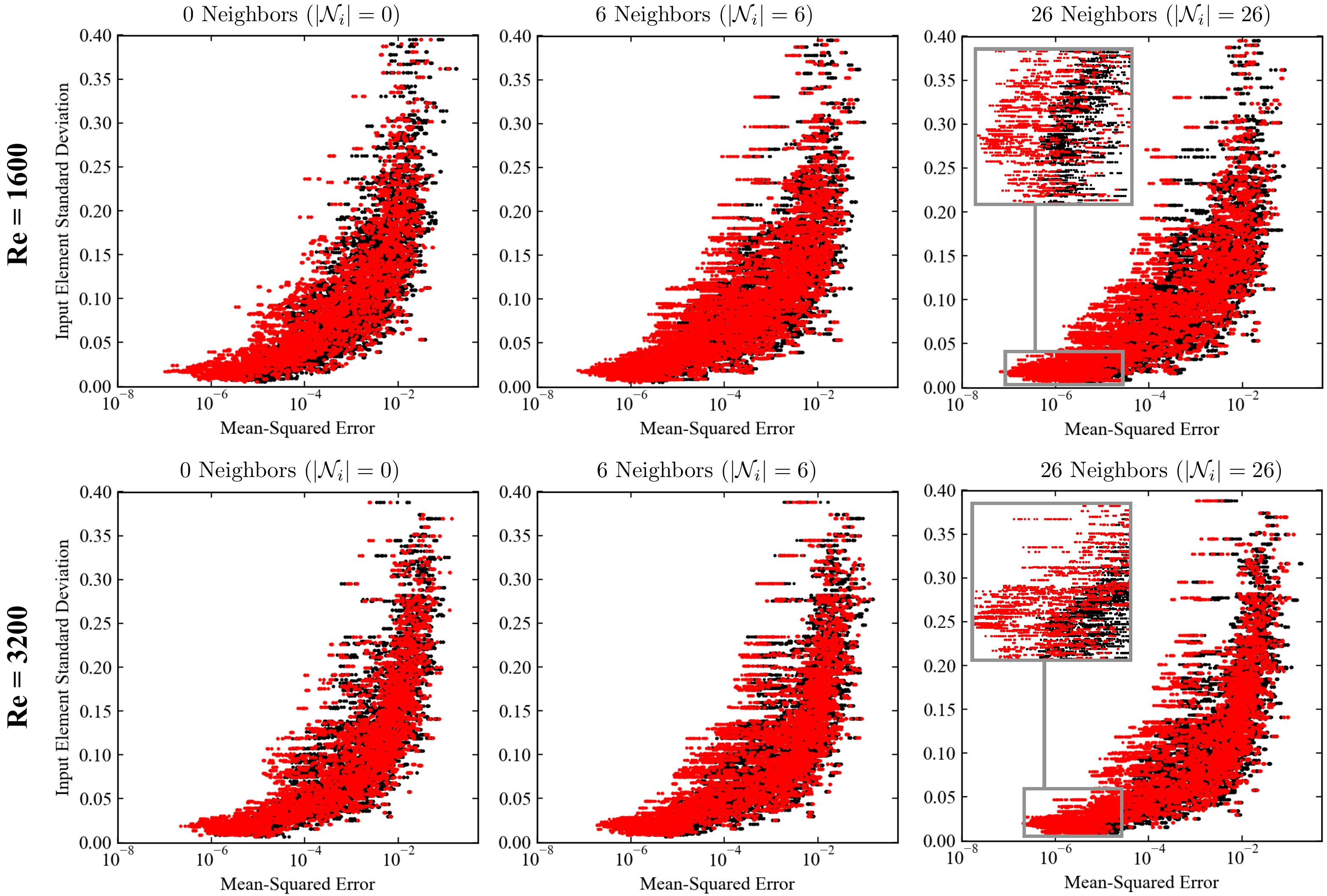}
    \caption{{\color{black}Standard deviation of input (coarse, P=1) element velocity fields versus mean-squared error in GNN predictions for 0 neighbors (left), 6 neighbors (middle), and 26 neighbors (right). Standard deviation and errors computed component-wise, with x-component results shown here (trends are consistent across other two components). Black markers denote Model 1 (coarse-scale GNN) predictions, and red markers denote Model 2 (multiscale GNN) predictions. Each marker corresponds to a single element in a test set snapshot. Top row shows Re=1600 predictions and bottom shows Re=3200.}}
    \label{fig:std_vs_error}
\end{figure}
}

\section{Proper Orthogonal Decomposition}
{\color{black}
\label{sec:app:pod}
Here, additional detail on the method of snapshots POD procedure \cite{sirovich1987} as it relates to the BFS spectrum analysis in Sec.~\ref{sec:results:bfs} is described. 

Ultimately, an orthonormal basis referred to as the \textit{POD basis} is recovered. This basis can be represented as the tall-and-skinny matrix ${\bf B} = [{\bf b}_1, \ldots, {\bf b}_{N_s}] \in \mathbb{R}^{3N_e N_p \times N_s}$, where $N_e$ is the number of elements, $N_p = 8^3$ is the number of discretization points-per-element in the P=7 case, and $N_s$ is the number of snapshots used to construct the basis (here, $N_s=20$). Since the full three-component velocity field is used in the POD, each vector ${\bf b}_i$ is the same size of a flattened three-component velocity field (of dimension $3N_e N_p$), and can therefore be reshaped to recover visualizations of the POD mode features corresponding to each velocity component. 

The instantaneous kinetic energy spectrum in the POD basis\footnote{Here, the POD basis vectors are normalized by the square root of the corresponding eigenvalue of the snapshot correlation matrix, such that ${\bf B}^{\text T}{\bf B} = {\bf I}$.} can be recovered as $\frac{1}{2}a_i(t)^2$, where $a_i(t)$ is the projection of some three-component fluctuating velocity field (such as one generated by a GNN reconstruction) onto the $i$-th POD basis vector. It is referred to as the scalar \textit{POD coefficient} corresponding to mode $i$, and is computed as
\begin{equation}
    a_i(t) = {\bf b}_i^{\text T} ({\bf Y}(t) - {\bf M}), \quad i=1,\ldots,N_b, 
\end{equation}
where ${\bf Y}(t)$ is a flattened velocity field snapshot at time $t$, and ${\bf M}$ is the flattened time-averaged velocity field sourced from the same set of P=7 target snapshots used to construct the POD basis ${\bf B}$. Before proceeding, the following points are emphasized: (1) the POD basis vectors are generated from \textit{target} (P=7) simulation snapshots, such that global errors can be accessed via deviations from the target POD energy spectrum, (2) separate POD bases are produced for Re=1600 and 3200, but the \textit{same} basis vectors are used to generate target and predicted spectra at a given Reynolds number, and (3) the POD carried out here serves as a mechanism to assess global errors, providing a pathway to evaluate GNN predictive strength relative to the target and the SE interpolation baseline, analogous to the conventional spectrum analysis approach used for the TGV. 
}

\section{Role of the Graph Unpooling Layer}
{\color{black}
\label{sec:app:unpool}
In summary, the unpooling layer is a graph node feature interpolation function, and allows the model to operate on complex-geometry representations (it is not constrained to ``regular" meshes or structured grids). Below are some additional details and explanation on the role of the unpooling operation in the context of the GNN super-resolution architecture. 

At a high level, the unpooling layer is used as a mechanism to interpolate the coarse node features, defined on the coarse P=1 query element, onto the fine P=7 element graph nodes. In the architectural configuration used here, the graph unpooling layer is used in two operational contexts: (1) as a residual connection \textit{outside} of the core GNN architecture, and (2) as an upsampling tool \textit{inside} of the GNN architecture. The former invocation of the unpooling layer (residual connection) takes as input the coarse velocity field on the query element: it forces the GNN to model the super-resolution task as a correction to the unpooling operation evaluated on the coarse element velocity field. The second invocation takes as input the hidden node features on the \textit{latent} query element (not the velocity field), and facilitates a multiscale GNN architecture; it serves as a connecting mechanism between coarse-scale and fine-scale message passing operations. As such, the interpolation is used twice for two different purposes, and the inputs to each layer call are fundamentally different.

To better understand the impact of the unpooling operation on the model, one can extract the interpolated velocity field due to the unpooling layer alone (\textit{without the added GNN residual correction}) and see how well it recovers the energy spectrum. This is shown in Fig.~\ref{fig:knn_interp}(a) for the TGV. Recall that, as described in Sec.~\ref{sec:unpool}, the graph unpooling layer leverages a K-nearest neighbors (KNN) algorithm to execute the interpolation, with K=8 coarse node neighbors. It can be seen that the KNN-based unpooling layer is, in some sense, recovering multiscale behavior as the energy is indeed activated at the higher wavenumbers (interestingly, there is more activation at the higher wavenumbers than the SE interpolation, and less at the lower wavenumbers). Also present is excessive overshoot at the high wavenumbers -- the hope is that, through the training procedure, the GNN learns to correct these overshoots through the residual correction model form. Observations of the corresponding GNN spectra throughout Sec.~\ref{sec:results:bfs} show that this is indeed the case. Additionally, although not shown here, it was found that models across the board converged to lower losses when including this unpooling layer via the residual connection. The increased performance comes at the added cost of executing an additional unpooling evaluation in the forward pass. 

\begin{figure}
    \centering
    \includegraphics[width=0.6\linewidth]{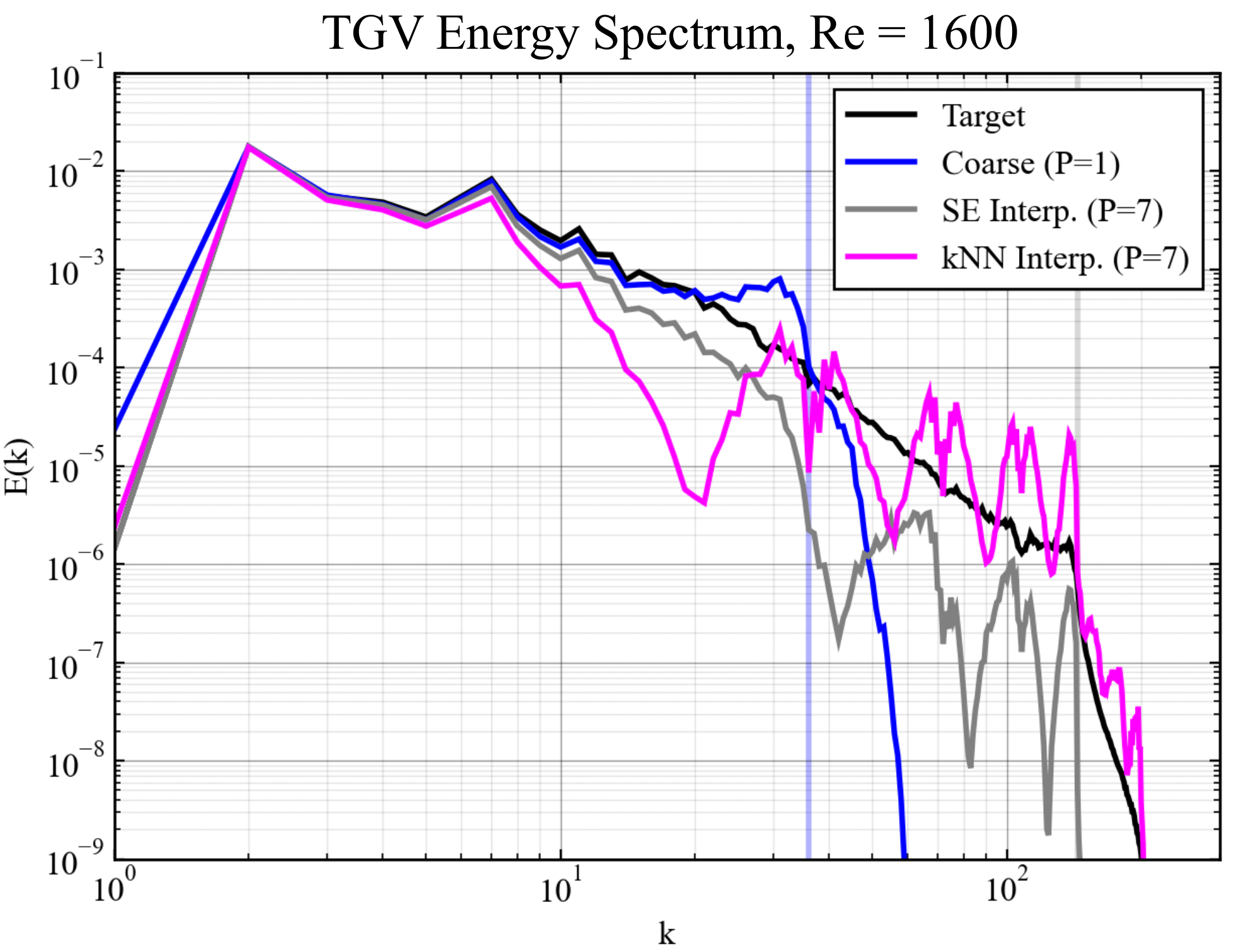}
    \caption{{\color{black}Instantaneous TGV energy spectrum at Re=1600, t=10.5, comparing target P=7 velocity field (black), coarse P=1 input field (blue), SE interpolation (gray), and KNN interpolation (magenta).}}
    \label{fig:knn_interp}
\end{figure}
}

\end{document}